\documentclass[a4paper,11pt]{article}
\pdfoutput=1

\usepackage{jheppub} 
\usepackage[T1]{fontenc} 
\usepackage{bm,latexsym,amsmath,amssymb,amsfonts,color}

\title{\boldmath Second-order effective energy-momentum tensor of gravitational scalar perturbations
with perfect fluid}

\author[a]{Inyong Cho}
\author[b]{Jinn-Ouk Gong}
\author[a,1]{Seung Hun Oh\note{Corresponding author}}

\affiliation[a]{School of Liberal Arts, Seoul National University of Science and Technology, Seoul 01811, Korea}
\affiliation[b]{Korea Astronomy and Space Science Institute, Daejeon 34055, Korea}

\emailAdd{iycho@seoultech.ac.kr}
\emailAdd{jgong@kasi.re.kr}
\emailAdd{shoh.physics@gmail.com}

\abstract{
We investigate the second-order gravitational scalar perturbations for a barotropic fluid.
We derive the effective energy-momentum tensor described by the quadratic terms of 
the gravitational and the matter perturbations. We show that the second-order effective 
energy-momentum tensor is gauge dependent. We impose three gauge conditions 
(longitudinal, spatially-flat, and comoving gauges) for dust and radiation. The resulting 
energy-momentum tensor is described only by a gauge invariant variable, but the 
functional form depends on the gauge choice. In the matter-dominated epoch with 
dust-like fluid background, the second-order effective energy density and pressure of 
the perturbations evolve as $1/a^2$ in all three gauge choices,
like the curvature density of the Universe, but they do not provide the correct equation of state.
The value of this parameter depends also on the gauge choice. In the 
radiation-dominated epoch, the perturbations in the short-wave limit behave in the same 
way as the radiation-like fluid in the longitudinal and the spatially-flat gauges. However, 
they behave in a different way in the comoving gauge. As a whole, we conclude that 
the second-order effective energy-momentum tensor of the scalar perturbation is strictly 
gauge dependent.
}

\begin{document}

\maketitle
\flushbottom

\section{Introduction}

The theory of cosmological perturbations is a great success in explaining the structure 
formation in the Universe. The density perturbation produced during inflation evolves 
in the subsequent Friedman universe, and produces the inhomogeneities in the matter 
distribution which can be observed in the cosmic microwave background and in the 
distribution of galaxies. The observation techniques have developed very rapidly, which 
allows us to investigate cosmology with a very high precision. Thus, now we are able to 
enter the regime of non-linearity with precise cosmological observations. This will reveal 
the new physics relevant for structure formation only accessible by non-linear 
cosmological perturbations.

The second-order cosmological perturbation with the scalar field in the inflation period 
was investigated in~\cite{Mukhanov:1996ak,Abramo:1997hu}. The authors investigated 
the second-order effective energy-momentum tensor (2EMT) of cosmological 
perturbations and its gauge invariance. Regarding the back-reaction effect of 2EMT in 
the Universe, for the long-wavelength perturbations, they found that the effective 
equation of state is given by $p_s \approx -\rho_s$ with $\rho_s<0$ for scalar 
perturbations, and $p_{gw} \approx -\rho_{gw}/3$ for tensor perturbations. More works 
have investigated the gauge invariance of the second-order perturbations 
in~\cite{Geshnizjani:2002wp,Brandenberger:2002sk,Geshnizjani:2003cn,Martineau:2005aa,Martineau:2005zu},
for example. However, the gauge invariance of 2EMT is 
questionable~\cite{Unruh:1998ic,Ishibashi:2005sj} and thus its physical relevance, 
under any approximation, is doubtful.

In this work, we will study the second-order cosmological {\it scalar} perturbations with 
{\it fluid} matter. From the Einstein's equation, we obtain 2EMT constructed by the quadratic 
combinations of scalar perturbations, and investigate its gauge invariance. Using the 
first-order equations, the matter perturbations can be expressed by the gravitational ones, 
in terms of which we can write 2EMT completely. The results show that 2EMT cannot be 
written only in terms of the gauge invariant variables such as the so-called Bardeen 
variable~\cite{Bardeen:1980kt}, but there remains gauge dependence. That is, for the 
cosmological scalar perturbations with fluid, the 2EMT is {\it not gauge invariant}.

More specifically, we investigate 2EMT in the Friedmann universe described by a 
barotropic fluid-like radiation and dust. Once we impose a specific gauge condition, 
e.g. longitudinal, spatially-flat, and comoving gauges, the resulting 2EMT can be 
expressed by the Bardeen variable only. However, the functional form of 2EMT varies 
depending on the gauge choice. Further we investigate some limits of this 2EMT 
depending on the wavelengths to see if the effect of 2EMT converges in all gauge choices.
In particular, we investigate the second-order effective equation of state evaluated from 
2EMT. The result shows that the values of the equation-of-state parameter are different 
for different gauge choices. This means that this effect of 2EMT is strictly gauge dependent,
so is not physically meaningful in any wavelength limit.

This paper is composed as following. In Section~\ref{secseEMT}, we introduce the 
gravitational scalar and the matter perturbations, derive 2EMT from the Einstein's equation,
and investigate its gauge invariance. In Section~\ref{sec3}, we investigate 2EMT in the 
Friedmann universe after imposing gauge conditions. In Section~\ref{secconc}, we conclude.

\section{Einstein's equation and 2EMT}
\label{secseEMT}

In this section, we derive the zeroth- and the first-order Einstein's equations. We also 
introduce gauge invariant variables, and write the first-order equation in a gauge 
invariant form. Then we construct 2EMT.

\subsection{Einstein Tensor}

Let us consider the cosmological perturbations in the Friedmann universe.
The most general metric perturbations are written as
\begin{align}
\label{metric1}
ds^{2} 
= a^{2}(\eta) \Big[ -(1+2A)d\eta^{2} - 2 B_{i} d\eta dx^{i}
+ (\delta_{ij} + 2 C_{ij}) dx^{i} dx^{j} \Big] 
\, ,
\end{align}
where the background metric represents a flat Friedmann universe, and the non-linear 
metric perturbations $A$, $B_i$ and $C_{ij}$ are both time- and space-dependent 
functions and are expanded in all orders, e.g. $A = A^{(1)} + A^{(2)} + A^{(3)} + \cdots$. 
With ${\cal H} \equiv a'/a=da/d\eta/a$, we can write each component of the Einstein 
tensor $G_{\mu\nu}$ up to second order in perturbations as follows~\cite{Gong:2017tev}:
\begin{align}
G_{00} 
= & 
3\mathcal{H}^{2} +2\mathcal{H}B_{k,k} +2\mathcal{H}C_{kk}^{\prime}    
-\Delta C_{kk} +C_{kl,kl} - (2\mathcal{H}^{\prime} +\mathcal{H}^{2})B_{k}B_{k}
+\frac{1}{2}B_{k,k}B_{l,l} 
\nonumber \\
& 
-\frac{1}{4}B_{k,l}B_{k,l} 
-\frac{1}{4}B_{k,l}B_{l,k} +B_{k}B_{l,lk} -B_{k}\Delta B_{k} 
-\frac{1}{2}C_{kl}^{\prime}C_{kl}^{\prime} + \frac{1}{2}C_{kk}^{\prime}C_{ll}^{\prime}
-4\mathcal{H}C_{kl}C_{kl}^{\prime} 
\nonumber \\ 
& 
+\frac{3}{2}C_{kl,m}C_{kl,m} 
-C_{kl,m}C_{km,l} -\frac{1}{2}(C_{kk,l} -2C_{lk,k})(C_{mm,l} -2C_{lm,m}) 
\nonumber \\
& 
+2C_{kl}(C_{mm,kl} +\Delta C_{kl} -2C_{mk,ml})     
-4\mathcal{H}A_{,k}B_{k} +2B_{k}(C_{mm,k}^{\prime} -C_{km,m}^{\prime})
\nonumber \\
& 
+2\mathcal{H}B_{k}(C_{mm,k} -2C_{km,m}) -B_{k,l}(C_{kl}^{\prime} +4\mathcal{H}C_{kl})   
+B_{k,k}C_{ll}^{\prime} -2A(\Delta C_{kk} -C_{kl,kl}) 
\, ,  
\label{ET1}  
\\
G_{0i} 
= & 
2\mathcal{H}A_{,i} +B_{[i,k]k} +(2\mathcal{H}^{\prime}
+ \mathcal{H}^{2})B_{i} +C_{ik,k}^{\prime}-C_{kk,i}^{\prime} 
-4\mathcal{H}AA_{,i} -B_{k}(B_{(i,k)}^{\prime} +2\mathcal{H}B_{[i,k]})   
\nonumber \\
& 
+B_{i}(B_{k,k}^{\prime} +2\mathcal{H}B_{k,k}) 
+(C_{mm,k} -2C_{km,m})C_{ik}^{\prime} +2C_{kl}(C_{kl,i}^{\prime} 
-C_{ik,l}^{\prime}) +C_{kl}^{\prime}C_{kl,i} 
\nonumber \\
& 
-2\mathcal{H}A^{\prime}B_{i} 
+A_{i,k}B_{k} +A_{,i}B_{k,k}
-A_{,k}B_{(i,k)} -2(2\mathcal{H}^{\prime} +\mathcal{H}^{2})AB_{i} 
\nonumber \\
& 
-(\Delta A)B_{i} -2\mathcal{H}B_{k}C_{ik}^{\prime} 
+2\mathcal{H}B_{i}C_{kk}^{\prime}       
-2B_{[i,k]l}C_{kl} +2B_{[i,k]}(C_{ll,k} -2C_{kl,l})
\nonumber \\
& 
-2B_{[k,l]}C_{i[k,l]}
-B_{i}(\Delta C_{kk} -C_{kl,kl}) +A_{,i}C_{kk}^{\prime}   
- A_{,k}C_{ik}^{\prime} 
\, ,   
\label{ET2}   
\end{align}
\begin{align}
G_{ij} 
= & 
G_{\textrm{D}} \delta_{ij} 
-A_{,ij} +B_{(i,j)}^{\prime} +2\mathcal{H}B_{(i,j)} +C_{ij}^{\prime \prime}
+2\mathcal{H}C_{ij}^{\prime} -\Delta C_{ij} -2(2\mathcal{H}^{\prime} 
+\mathcal{H}^{2}) C_{ij} +2C_{k(i,j)k} 
\nonumber \\
& 
-C_{kk,ij}  +A_{,i}A_{,j} +2AA_{,ij} +B_{k,k}B_{(i,j)} +B_{k}B_{(i,j)k}
-\frac{1}{2}B_{i,k}B_{j,k} -\frac{1}{2}B_{k,i}B_{k,j} 
\nonumber \\
& 
-B_{k}B_{k,ij}  -2C_{kk}^{\prime \prime} C_{ij}     
+C_{kk}^{\prime}C_{ij}^{\prime} 
-2C_{ik}^{\prime}C_{jk}^{\prime} -4\mathcal{H}C_{kk}^{\prime}C_{ij} 
+2(\Delta C_{kk} -C_{kl,kl})C_{ij}
\nonumber \\
& 
+2C_{kl}(C_{ij,kl} +C_{kl,ij} -2C_{k(i,j)l}) 
-(C_{kk,l}-2C_{lk,k})(C_{ij,l} -2C_{l(i,j)}) +C_{kl,i}C_{kl,j} 
\nonumber \\
& 
+2C_{ik,l}(C_{jk,l} -C_{jl,k}) -A^{\prime}B_{(i,j)}  
-2A(B_{(i,j)}^{\prime} +2\mathcal{H}B_{(i,j)}) -2B_{k}(C_{k(i,j)}^{\prime}
-C_{ij,k}^{\prime}) 
\nonumber \\
& 
-(B_{k}^{\prime} +2\mathcal{H}B_{k})(2C_{k(i,j)} -C_{ij,k})      
+B_{k,k}(C_{ij}^{\prime} -4\mathcal{H}C_{ij}) -2B_{k,k}^{\prime}C_{ij}
+B_{(i,j)}C_{kk}^{\prime}
\nonumber \\
& 
-B_{i,k}C_{jk}^{\prime} -B_{j,k}C_{ik}^{\prime} -2A(C_{ij}^{\prime \prime} 
+2\mathcal{H}C_{ij}^{\prime})    
-A^{\prime}(C_{ij}^{\prime} -4\mathcal{H}C_{ij}) +4(2\mathcal{H}^{\prime} 
+\mathcal{H}^{2})AC_{ij} 
\nonumber \\
& 
+A_{,k}(2C_{k(i,j)} -C_{ij,k}) +2(\Delta A)C_{ij} 
\, ,
\label{ET3} 
\end{align}
where $G_{\textrm{D}}$ multiplied by the Kronecker delta $\delta_{ij}$ in $G_{ij}$ is given by
\begin{align}
G_{\textrm{D}} = 
& 
-(2\mathcal{H}^{\prime} +\mathcal{H}^{2}) 
+2\mathcal{H}A^{\prime} +2(2\mathcal{H}^{\prime} +\mathcal{H}^{2})A +\Delta A
-B_{k,k}^{\prime} -2\mathcal{H}B_{k,k} -C_{kk}^{\prime\prime} 
-2\mathcal{H}C_{kk}^{\prime}    
\nonumber \\
& 
+\Delta C_{kk} -C_{kl,kl} -8\mathcal{H}AA^{\prime} -(\nabla A)^{2} -2A\Delta A
-4(2\mathcal{H}^{\prime} +\mathcal{H}^{2})A^{2}
+2\mathcal{H}B_{k}B_{k}^{\prime} 
\nonumber \\
& 
+(2\mathcal{H}^{\prime} +\mathcal{H}^{2})B_{k}B_{k} +\frac{3}{4}B_{k,l}B_{k,l} 
-\frac{1}{4}B_{k,l}B_{l,k} -\frac{1}{2}B_{k,k}B_{l,l} -B_{l}B_{k,kl} 
+B_{k} \Delta B_{k}  
\nonumber \\
&
+2C_{kl}C_{kl}^{\prime \prime} +\frac{3}{2}C_{kl}^{\prime}C_{kl}^{\prime}
-\frac{1}{2}C_{kk}^{\prime}C_{ll}^{\prime} +4\mathcal{H}C_{kl}C_{kl}^{\prime} 
+\frac{1}{2}(C_{kk,l} -2C_{lk,k})(C_{mm,l} -2C_{lm,m})   
\nonumber \\
& 
-\frac{3}{2}C_{lm,k}C_{lm,k} +C_{lm,k}C_{lk,m}
-2C_{kl}(C_{mm,kl} +\Delta C_{kl} -2C_{mk,ml}) +A^{\prime}B_{k,k} 
+2AB_{k,k}^{\prime}          
\nonumber \\
& 
+2\mathcal{H}A_{,k}B_{k} +4\mathcal{H}AB_{k,k} +2B_{k,l}^{\prime}C_{kl}
-B_{k}^{\prime}(C_{ll,k}-2C_{kl,l}) -2B_{k}(C_{ll,k}^{\prime}-C_{kl,l}^{\prime})
\nonumber \\
&
-B_{k,k}C_{ll}^{\prime}   
+B_{k,l}C_{kl}^{\prime} +4\mathcal{H}B_{k,l}C_{kl} 
-2\mathcal{H}B_{l}(C_{kk,l} -2C_{lk,k}) 
+2AC_{kk}^{\prime\prime} +A^{\prime}C_{kk}^{\prime}    
\nonumber \\ 
&
+4\mathcal{H}AC_{kk}^{\prime}           
-2A_{,kl}C_{kl}  +A_{,k}(C_{ll,k} -2C_{kl,l})
\, .
\end{align}

\subsection{Matter Energy-Momentum Tensor}

Now we consider a perfect fluid matter of which the energy-momentum tensor is given by
\begin{equation}
\label{Tpf}
T_{\mu \nu} 
= (\rho +p) u_{\mu}u_{\nu} +pg_{\mu \nu} 
\, ,,
\end{equation}
where $\rho$ is the energy density, $p$ is the pressure, and $u^\mu$ is the four-velocity. 
The perturbations of fluid are written as
\begin{equation}
\rho = \rho_{0} + \delta \rho 
\quad \text{and} \quad
p = p_{0} + \delta p 
\, ,
\end{equation}
where the subscript $0$ denotes the background unperturbed quantity. The perturbations 
are, as those in the metric, expanded in all orders, for example, 
$\delta\rho = \delta\rho^{(1)} + \delta\rho^{(2)} + \delta\rho^{(3)} + \cdots$. 
If we introduce the peculiar velocity $v^{i}$ defined by~\cite{Kodama:1985bj}
\begin{equation}
v^{i} \equiv \frac{dx^{i}}{dx^{0}} = \frac{u^{i}}{u^{0}} , 
\end{equation}
we can find up to second order in perturbations, with the normalization condition 
$u^{\mu}u_{\mu} = - 1$,
\begin{align}
u^{0} 
& = 
\frac{1}{a} \Big( 1-A +\frac{3}{2}A^{2} -B_{k}v_{k} + \frac{1}{2}v_{k}v_{k}  \Big) 
\, ,
\\
u^{i} 
& = 
\frac{1}{a} (1-A)v^{i}  
\, .
\end{align}
Then each component of $T_{\mu\nu}$ is obtained as, up to second order in perturbations,
\begin{align}
T_{00} 
= & 
a^{2} \big[ \rho_{0} +\delta \rho +2A\rho_{0}
+2A\delta \rho +(\rho_{0} +p_{0})v_{k}v_{k} \big]  
\, ,  
\label{pfemt1} 
\\
T_{0i} 
= & 
a^{2} \big[ \rho_{0}(B_{i} -v_{i}) -p_{0} v_{i}
+\delta \rho (B_{i}-v_{i}) - \delta p v_{i}  -2(\rho_{0} +p_{0}) C_{ik} v_{k} 
\big] 
\, ,     
\label{pfemt2}   
\\
T_{ij} 
= & 
a^{2} \big[ p_{0} \delta_{ij} +2 p_{0} C_{ij} 
+ \delta p \delta_{ij} +2C_{ij} \delta p + (\rho_{0} +p_{0})(B_{i} -v_{i})
(B_{j} -v_{j})      \big]    
\, .
\label{pfemt3}  
\end{align} 
%
%In this work, we shall consider only upto the second order.
%We assume that the linear perturbations in the 2nd order 
%[e.g., $\delta g_{\mu\nu}^{(2)}$,  $\delta\rho^{(2)}$, and $\delta p^{(2)}$] vanish
%after the temporal integration due the their stochastic behaviour.
%Therefore, the perturbation functions shown in Eqs.~\eqref{pfemt1}-\eqref{pfemt3}
%represent the first order only.
%
In this work, we shall consider only a perfect fluid matter.

\subsection{Einstein's Equation}

The Einstein's equation order by order can be constructed by equating the Einstein tensor 
in Eqs.~\eqref{ET1}-\eqref{ET3} and the corresponding matter energy-momentum tensor in
Eqs.~\eqref{pfemt1}-\eqref{pfemt3}.

\subsubsection{Zeroth-order equations}

The zeroth-order, background equations are given by the $00$ and $ij$ components: 
\begin{align}
3\mathcal{H}^{2} 
& = 
8\pi G a^{2} \rho_{0}  
\,  ,   
\label{pf0th1}   
\\
2\mathcal{H}^{\prime} +\mathcal{H}^{2} 
& = 
-8 \pi G a^{2} p_{0}      
\, .
\label{pf0th2}
\end{align}
Note that combining these two equations gives the conservation equation for the background 
energy density $\rho_0$, which can be also derived from the conservation of energy-momentum 
tensor:
\begin{equation}
\rho_{0}^{\prime} + 3\mathcal{H} (\rho_{0} + p_{0}) = 0  \, .
\label{pf0thmatter}
\end{equation}

\subsubsection{First-order equations and gauge invariant variables}

Using the background equations, we can cast the first-order Einstein's equation into the following:
\begin{align}
&
6 \mathcal{H}^{2} A  - 2\mathcal{H} B_{k,k} 
- 2 \mathcal{H} C^{\prime}_{kk} + \Delta C_{kk} -C_{kl,kl} 
 = 
- 8\pi G a^{2} \delta \rho  
\, ,   
\label{pf1st1}   
\\
&
2 \mathcal{H} A_{,i} + B_{[i,k]k} + 2(\mathcal{H}^{\prime} -\mathcal{H}^{2}) B_{i}
+ C^{\prime}_{ik,k} - C^{\prime}_{kk,i}
 = 
2(\mathcal{H}^{\prime} -\mathcal{H}^{2}) v_{i} 
\, ,   
\label{pf1st2}  
\\
&
\delta_{ij} \big[ 2\mathcal{H}A^{\prime} 
+ 2(2 \mathcal{H}^{\prime} +  \mathcal{H}^{2})A + \Delta A - B_{k,k}^{\prime} 
-2\mathcal{H}B_{k,k} - C_{kk}^{\prime \prime} -2\mathcal{H} C_{kk}^{\prime}
+\Delta C_{kk}  - C_{kl,kl} \big] 
\notag\\ 
&
-A_{,ij} + B^{\prime}_{(i,j)} +2\mathcal{H}B_{(i,j)} +C_{ij}^{\prime\prime} 
+ 2\mathcal{H} C_{ij}^{\prime}-\Delta C_{ij} + 2C_{k(i,j)k} - C_{kk,ij} 
= 
8\pi G a^{2} \delta p \ \delta_{ij} 
\, .   
\label{pf1st3}
\end{align}
The first-order conservation equations read
\begin{align}
\delta \rho^{\prime} + 3\mathcal{H} (\delta \rho + \delta p ) 
+ (C_{kk}^{\prime} + v_{k,k} ) (\rho_{0} + p_{0}) 
& = 0  
\, ,   
\label{pf1stmatter1} 
\\
\delta p_{,i} - (B_{i} - v_{i}) p_{0}^{\prime} 
+ [ A_{,i} -(B_{i}^{\prime} - v_{i}^{\prime}) 
-\mathcal{H} (B_{i} - v_{i})]  (\rho_{0} + p_{0}) 
& = 0 
\, . 
\label{pf1stmatter2}
\end{align}

Now let us consider the gauge transformation of cosmological perturbations.
As is well known, not all degrees of freedom in cosmological perturbations are
physical. This is because a generic coordinate transformation should not modify
the physics of perturbations. But in general cosmological perturbations are
subject to the change of coordinates: Let a function $Q$ be any scalar, vector or 
tensor quantity. Then, under the infinitesimal coordinate transformation
$x^\mu \to x^\mu + \xi^\mu$, the corresponding gauge transformation for $Q$ can 
be written as
\begin{equation}
Q \to Q - \mathcal{L}_\xi Q
\, ,
\end{equation}
where $\mathcal{L}_\xi$ denotes the Lie derivative in the direction of $\xi^\mu$.
Then the metric perturbations in \eqref{metric1} transform as
\begin{align}
A & \to A - {\xi^0}' - \mathcal{H}\xi^0
\, ,
\\
B_i & \to B_i - \xi^0{}_{,i} + \xi_i'
\, ,
\\
C_{ij} & \to C_{ij} - \mathcal{H}\xi^0\delta_{ij} - \xi_{(i,j)}
\, .
\end{align}
If we consider only scalar perturbations such that
\begin{equation}
\label{ABC}
A = \alpha \, , 
\quad
B_i = \beta_{,i} \, , 
\quad
C_{ij} = -\psi\delta_{ij} + E_{,ij} 
\, ,
\end{equation}
they transform as follows~\cite{Noh:2004bc}:
\begin{align}
\alpha & \to \alpha - {\xi^0}' - \mathcal{H}\xi^0
\, ,
\\
\beta & \to \beta - \xi^0 + \xi'
\, ,
\\
\psi & \to \psi + \mathcal{H}\xi^0
\, ,
\\
E & \to E - \xi 
\, ,
\end{align}
where $\xi_i = \xi_{,i}$.
Then, as in Ref.~\cite{Bardeen:1980kt}, one choice of gauge invariant variables is the following:
\begin{align}
\Phi &= \alpha - Q^{\prime} - \mathcal{H} Q \, , 
\label{Phi}
\\
\Psi &= \psi + \mathcal{H} Q \, , 
\label{Psi}
\\
\overline{\delta \rho} &= \delta \rho - \rho_{0}^{\prime} Q \, ,  
\\
\overline{\delta p} &= \delta p - p_{0}^{\prime} Q \, , 
\\
\overline{v}_{i} &= v_{i} + E_{,i}^{\prime} \, ,   
\end{align}
where 
\begin{equation}
Q = \beta + E^{\prime} \, .
\end{equation}
From  the off-diagonal components of Eq.~\eqref{pf1st3}, we have $\Phi = \Psi$. This 
means that the off-diagonal component of $T_{ij}$ vanishes. Then the first-order 
equations \eqref{pf1st1}-\eqref{pf1st3} can be written in a gauge invariant form as
\begin{align}
3\mathcal{H} \Psi^{\prime} + 3 \mathcal{H}^{2} \Psi - \Delta \Psi 
& = 
- 4 \pi G a^{2} \overline{\delta \rho}  
\, , 
\label{gi1sthyd1} 
\\
\big(  \Psi^{\prime} + \mathcal{H} \Psi \big)_{,i} 
& = 
(  \mathcal{H}^{\prime} - \mathcal{H}^{2} ) \ \overline{v}_{i} 
\, , 
\label{gi1sthyd2} 
\\
\Psi^{\prime \prime} + 3\mathcal{H} \Psi^{\prime} 
+ (2 \mathcal{H}^{\prime} +  \mathcal{H}^{2} ) \Psi
& = 
4 \pi G a^{2} \overline{\delta p} 
\, .  
\label{gi1sthyd3} 
\end{align}

\subsubsection{Second-order effective energy-momentum tensor}
\label{sectau}

The second-order effective energy-momentum tensor (2EMT) is constructed in the following way
\cite{Mukhanov:1996ak,Abramo:1997hu,Ishibashi:2005sj}. Let us consider the Einstein's equation
at second-order in perturbations:
\begin{align}
G^{(2)}_{\mu\nu} = 8\pi G T^{(2)}_{\mu\nu} \, . 
\end{align}
The second-order Einstein tensor consists of two parts, 
\begin{align}
\label{eq:G2-2parts}
G^{(2)}_{\mu\nu} = G^{(1)}_{\mu\nu}[g^{(2)}] + G^{(2)}_{\mu\nu}[g^{(1)}] \, ,
\end{align}
where $G^{(1)}_{\mu\nu}[g^{(2)}]$ represents the linear terms in genuine second-order perturbations,
and $G^{(2)}_{\mu\nu}[g^{(1)}]$ represents the quadratic terms in first-order perturbations.
The former describes the  second-order part of the geometry, while the latter contributes as a part of 
the second-order energy-momentum tensor. $T^{(2)}_{\mu\nu}$ can also be decomposed into two parts  
as \eqref{eq:G2-2parts}, and the second-order effective energy-momentum tensor is constructed as
\begin{align}
G^{(1)}_{\mu\nu}[g^{(2)}] =   
8\pi G T^{(1)}_{\mu\nu}[g^{(2)},\delta\rho^{(2)},\delta p^{(2)}]
+ \underbrace{8\pi G T^{(2)}_{\mu\nu}[g^{(1)},\delta\rho^{(1)},\delta p^{(1)}]
- G^{(2)}_{\mu\nu}[g^{(1)}]}_{
\equiv  8\pi G T^{(2,\rm{eff})}_{\mu\nu}}
\, .
\end{align}
Using the first-order Eqs.~\eqref{pf1st1}-\eqref{pf1st3} [or \eqref{gi1sthyd1}-\eqref{gi1sthyd3}],
we can replace the matter perturbations with the metric ones in $T^{(2)}_{\mu\nu}$.
Then finally we can obtain $T^{(2,\rm{eff})}_{\mu\nu}$ in terms of the Bardeen variable $\Psi$ and
the gauge dependent variables $Q = \beta + E^{\prime}$ and $E$. After integrating over several 
wavelengths, which is denoted by braket notations, the total derivative terms are integrated out,
and finally we find $\tau_{\mu\nu} \equiv \left\langle T^{(2,\rm{eff})}_{\mu\nu} \right\rangle$ as
\begin{align}
\tau_{00} 
= & 
\frac{1}{8\pi G } \bigg[ 
-\frac{2}{\mathcal{H}^{\prime} -\mathcal{H}^{2} }\big\langle (\nabla \Psi^{\prime})^{2} \big\rangle 
- \frac{4\mathcal{H}}{\mathcal{H}^{\prime} -\mathcal{H}^{2} }\big\langle \nabla \Psi^{\prime}
\cdot \nabla \Psi \big\rangle 
+\frac{5\mathcal{H}^{\prime} -7\mathcal{H}^{2} }{\mathcal{H}^{\prime} -\mathcal{H}^{2}} 
\big\langle (\nabla \Psi)^{2} \big\rangle          
\nonumber \\
& \qquad
- 3 \big\langle (\Psi^{\prime})^{2} \big\rangle    
-12\mathcal{H}^{2} \big\langle \Psi^{2} \big\rangle
+6 \big\langle (\mathcal{H}^{\prime} Q - \mathcal{H}Q^{\prime} -4\mathcal{H}^{2} Q) 
\Psi^{\prime}  \big\rangle 
+ 2 \big\langle \nabla Q \cdot \nabla \Psi^{\prime} \big\rangle
\nonumber \\ 
& \qquad
- 4 \mathcal{H} \big\langle \nabla Q \cdot \nabla \Psi \big\rangle  
-24 \mathcal{H}^{2} \big\langle (Q^{\prime} + \mathcal{H} Q)\Psi  \big\rangle
-2\mathcal{H} \big\langle \nabla Q^{\prime} \cdot \nabla Q  \big\rangle 
-3\big\langle (\mathcal{H}^{\prime} Q - \mathcal{H}Q^{\prime})^{2}  \big\rangle
\nonumber \\
& \qquad
+12\mathcal{H}^{2}(2\mathcal{H}^{\prime} -\mathcal{H}^{2})\big\langle Q^{2} \big\rangle   
+ \big\langle \Delta E^{\prime}
\big( 2 \mathcal{H} \Psi + 4\mathcal{H}^{2} Q - 3 \mathcal{H}^{2} E^{\prime}
+ 2\mathcal{H} \Delta E  \big)  \big\rangle       
\nonumber \\
& \qquad 
-2 \big\langle \Delta E
\big( 2 \mathcal{H} \Psi^{\prime} - \Delta \Psi -2\mathcal{H}^{2} Q^{\prime} 
- 2 \mathcal{H} \mathcal{H}^{\prime} Q   \big)  \big\rangle
\bigg]    
\, ,    
\label{gipf1} 
\\ 
\tau_{0i} = & 0 \, ,  
\label{gipf2}  
\\
\tau_{ij}  
= &  \frac{1}{8\pi G}\delta_{ij}
\bigg\{ -\frac{2}{3(\mathcal{H}^{\prime} -\mathcal{H}^{2})}\big\langle (\nabla \Psi^{\prime})^{2} \big\rangle 
- \frac{4\mathcal{H}}{3(\mathcal{H}^{\prime} -\mathcal{H}^{2})}
\big\langle \nabla \Psi^{\prime} \cdot \nabla \Psi \big\rangle  
-  \frac{\mathcal{H}^{\prime} +\mathcal{H}^{2}}{3(\mathcal{H}^{\prime} -\mathcal{H}^{2})} \big\langle (\nabla \Psi)^{2} \big\rangle 
\nonumber \\
& \hspace{3em}
+ \big\langle (\Psi^{\prime})^{2} \big\rangle   
+8 \mathcal{H}  \big\langle \Psi \Psi^{\prime} \big\rangle
+4(2\mathcal{H}^{\prime} +\mathcal{H}^{2}) \big\langle \Psi^{2} \big\rangle    
+4\big\langle (Q^{\prime} + 2 \mathcal{H} Q) \Psi^{\prime\prime} \big\rangle    
\nonumber \\
& \hspace{3em}
+ 2 \big\langle \big\{  Q^{\prime \prime} + 9\mathcal{H} Q^{\prime} 
+ (\mathcal{H}^{\prime} +12\mathcal{H}^{2})Q\big\} \Psi^{\prime} \big\rangle    
\nonumber \\
& \hspace{3em}
+4 \big\langle \big\{ 2\mathcal{H} Q^{\prime \prime} + 4(\mathcal{H}^{\prime} +\mathcal{H}^{2}) 
Q^{\prime} + 2\mathcal{H} (3\mathcal{H}^{\prime} +\mathcal{H}^{2})Q\big\} 
\Psi\big\rangle   
+ \frac{2}{3} \big\langle \nabla Q^{\prime \prime} \cdot \nabla Q \big\rangle 
\nonumber \\
& \hspace{3em}
+ \frac{2}{3} \big\langle  (\nabla Q^{\prime})^{2} \big\rangle  
+ \frac{4}{3}\mathcal{H}  \big\langle \nabla Q^{\prime} \cdot \nabla Q \big\rangle  
+ 2\mathcal{H} \big\langle Q^{\prime \prime} Q^{\prime}  \big\rangle
- 2\mathcal{H}^{\prime} \big\langle Q^{\prime \prime} Q \big\rangle  
+ \mathcal{H}^{2}  \big\langle Q^{\prime 2}  \big\rangle
\nonumber \\
& \hspace{3em}
- 2(2\mathcal{H}^{\prime \prime} + 3\mathcal{H} \mathcal{H}^{\prime}) 
\big\langle Q^{\prime} Q  \big\rangle     
- (8\mathcal{H} \mathcal{H}^{\prime \prime} +3\mathcal{H}^{\prime 2} 
- 4\mathcal{H}^{4})\big\langle Q^{2}  \big\rangle   
\nonumber \\
& \hspace{3em}
- \frac{2}{3} \big\langle \Delta E^{\prime \prime}
\big(\Psi +2\mathcal{H} Q - 3\mathcal{H}E^{\prime}+2\Delta E \big)\big\rangle   
\nonumber \\
& \hspace{3em}
- \frac{1}{3} \Big\langle \Delta E^{\prime}
\big[ 4\Psi^{\prime} + 10\mathcal{H} \Psi + 8\mathcal{H}Q^{\prime}  
+ 8 (\mathcal{H}^{\prime} + \mathcal{H}^{2}) Q
-3(2\mathcal{H}^{\prime} +\mathcal{H}^{2}) E^{\prime}
\nonumber \\
& \hspace{8em}
+ 2\Delta E^{\prime} + 8\mathcal{H} \Delta E \big] \Big\rangle  
\nonumber \\
& \hspace{3em}
+ \frac{2}{3} \Big\langle \Delta E
\big[ 2\Psi^{\prime \prime} + 4\mathcal{H} \Psi^{\prime} 
-2\mathcal{H}Q^{\prime \prime}  
-4 (\mathcal{H}^{\prime} + \mathcal{H}^{2}) Q^{\prime}
-2 (\mathcal{H}^{\prime \prime} + 2\mathcal{H}^{\prime}\mathcal{H}  )Q
\nonumber \\
& \hspace{8em}
+ \Delta ( E^{\prime \prime} + 2\mathcal{H} E^{\prime} ) \big] \Big\rangle 
\bigg\}     
\, .  
\label{gipf3}
\end{align} 
This energy-momentum tensor $\tau_{\mu\nu}$ is the explicit expression of the 2EMT of the scalar 
perturbations for perfect fluid. As we can see, the explicit form of $\tau_{\mu\nu}$ is expressed
not only by the gauge invariant variable $\Psi$, but also by the gauge dependent variables $Q$ and $E$.
Therefore, we can conclude that  2EMT is definitely gauge dependent.

\section{2EMT in Friedmann universe: gauge choices}
\label{sec3}

In the previous section, we have obtained the explicit form of 2EMT $\tau_{\mu\nu}$ 
and have noticed that it is gauge dependent. But, nevertheless, we may expect that
under different gauge conditions we still obtain similar behavior of the equation of state
in certain wavelength limits so that the explicit gauge dependence does not matter
practically.
In this section, we investigate the 2EMT in the Friedmann universe, in particular, 
during the radiation-dominated epoch (RDE) and the matter-dominated epoch (MDE).
The corresponding background equation of state $w\equiv p_0/\rho_0$ is $1/3$ and $0$ 
respectively. We evaluate the 2EMT in three gauge choices -- longitudinal, spatially-flat, 
and comoving gauges.

For isentropic process ($\delta S=0$), using the relation
\begin{align}
\delta p 
= 
\left( \frac{\partial p}{\partial\rho} \right)_{S} \delta\rho
+ \left( \frac{\partial p}{\partial S} \right)_{\rho} \delta S 
\equiv 
c_s^2 \delta\rho
\, ,
\end{align}
where  $c_s^2 = w$ for a barotropic fluid, the second-order differential equation for the 
Bardeen variable $\Psi$ is obtained from Eqs.~\eqref{gi1sthyd1} and \eqref{gi1sthyd3}:
\begin{align}
\label{Psieq}
\Psi'' + 3(1+w){\cal H}\Psi' -w\Delta\Psi +[2{\cal H}' + (1+3w){\cal H}^2]\Psi= 0 \, .
\end{align}
With the Fourier-mode expansion
\begin{align}
\Psi (\eta,{\bm x}) = \sum_{{\bm k}}  {\Psi}_{{\bm k}}(\eta) e^{i {\bm k} \cdot {\bm x}}
\, , 
\end{align}
we find the solution for MDE ($w=0$) and RDE ($w=1/3$) as~\cite{Mukhanov:2005sc}
\begin{align}
\label{eq:Psi-sol}
\Psi_{\bm k}(\eta) = \left\{
\begin{array}{ll}
{c}_{1}(k) + \dfrac{{c}_{2}(k)}{\eta^{5}} & (\text{MDE})
\vspace{0.5em}
\\
\dfrac{ {d}_{1}(k) }{\eta^{3}}
\left[ \dfrac{k\eta}{\sqrt{3}} \cos \left( \dfrac{k\eta}{\sqrt{3}} \right) - \sin \left( \dfrac{k\eta}{\sqrt{3}} \right) \right]  
+\dfrac{ {d}_{2}(k) }{\eta^{3}}  
\left[ \dfrac{k\eta}{\sqrt{3}} \sin \left( \dfrac{k\eta}{\sqrt{3}} \right) + \cos \left( \dfrac{k\eta}{\sqrt{3}} \right) \right] 
& (\text{RDE})
\end{array}
\right.
\, ,
\end{align}
where $c_1(k)$, $c_2(k)$, $d_1(k)$ and $d_2(k)$ are momentum-dependent functions constant in time, 
and $1/\sqrt{3}$ is the sound speed during RDE.
Now we fix the gauge condition, obtain $\tau_{\mu\nu}$, and plug in the above solutions to see 
the behaviour of 2EMT in the Friedmann universe. We interpret the components of  2EMT 
as the second-order effective energy density and pressure,
$\tau^\mu{}_\nu = \text{diag}(-\varrho,\mathfrak{p},\mathfrak{p},\mathfrak{p})$.
We present only the dominant terms in $\tau_{\mu\nu}$ here, with more complete results 
being given in Appendix~\ref{app:2emt-mdrd}.

\subsection{Longitudinal gauge}

Let us take the longitudinal gauge by imposing the conditions,
\begin{align} 
\label{LG}
\beta = E= 0 \, , 
\end{align}
which gives $Q = 0$.
As the gauge variables $E$ and $Q$ in Eqs.~\eqref{gipf1}-\eqref{gipf3} vanish,
$\tau_{\mu\nu}$ is expressed only by the Bardeen variable $\Psi$ as
\begin{align}
\tau_{00} 
= & 
\frac{1}{8\pi G } \bigg[ 
-\frac{2}{\mathcal{H}^{\prime} -\mathcal{H}^{2} }\big\langle (\nabla \Psi^{\prime})^{2} \big\rangle 
- \frac{4\mathcal{H}}{\mathcal{H}^{\prime} -\mathcal{H}^{2} }\big\langle \nabla \Psi^{\prime}
\cdot \nabla \Psi \big\rangle 
+\frac{5\mathcal{H}^{\prime} -7\mathcal{H}^{2} }{\mathcal{H}^{\prime} -\mathcal{H}^{2}} 
\big\langle (\nabla \Psi)^{2} \big\rangle            
\nonumber \\
& \hspace{3em}
- 3 \big\langle (\Psi^{\prime})^{2} \big\rangle 
-12\mathcal{H}^{2} \big\langle \Psi^{2} \big\rangle \bigg]   
\, ,    
\label{pfeffemlongM1}   
\\
\tau_{ij} 
= & 
\frac{1}{8\pi G  } \delta_{ij}
\bigg[ -\frac{2}{3(\mathcal{H}^{\prime} -\mathcal{H}^{2})}\big\langle (\nabla \Psi^{\prime})^{2} \big\rangle 
- \frac{4\mathcal{H}}{3(\mathcal{H}^{\prime} -\mathcal{H}^{2})}
\big\langle \nabla \Psi^{\prime} \cdot \nabla \Psi \big\rangle  
-  \frac{\mathcal{H}^{\prime} +\mathcal{H}^{2}}{3(\mathcal{H}^{\prime} -\mathcal{H}^{2})}
\big\langle (\nabla \Psi)^{2} \big\rangle   
\nonumber \\
& \hspace{3em}
+8 \mathcal{H}  \big\langle \Psi \Psi^{\prime} \big\rangle + \big\langle (\Psi^{\prime})^{2} \big\rangle
+4(2\mathcal{H}^{\prime} +\mathcal{H}^{2}) \big\langle \Psi^{2} \big\rangle  \bigg]     
\, .
\label{pfeffemlongM3}
\end{align}

\subsubsection{MDE}

Plugging the solution \eqref{eq:Psi-sol} in Eqs.~\eqref{pfeffemlongM1} and \eqref{pfeffemlongM3}, 
we find the dominant terms as
\begin{align}
\tau_{00} 
&= 
\frac{1}{8\pi G} \bigg[ \frac{19k^{2}}{3} |{c}_{1}|^{2}  
+ \mathcal{O} \left( \eta^{-2} \right) \bigg]
\, ,
\\
\tau_{ij}
& = 
\frac{1}{8\pi G} \delta_{ij} \bigg[ \frac{k^{2}}{9} |{c}_{1}|^{2}  
+ \mathcal{O} \left( \eta^{-5} \right) \bigg]
\, .
\end{align}
Recalling that $a(\eta) \propto \eta^2$ during MDE, for the most dominant terms, we have
\begin{align}
\label{eosa2LG}
\varrho = \frac{\tau_{00}}{a^2} \propto \frac{1}{a^2}
\quad \text{and} \quad
\mathfrak{p} = \frac{\tau_{ii}}{a^2} \propto \frac{1}{a^2}
\, ,
\end{align}
which individually exhibit the same $a$-dependence as the curvature density. 
But their relation, $\mathfrak{p} \approx 57\varrho$, is not that of the curvature density.

\subsubsection{RDE}

Using the solution \eqref{eq:Psi-sol} for \eqref{pfeffemlongM1} and \eqref{pfeffemlongM3}, 
we find very complicated results containing a lot of sine and cosine functions.
To simplify our discussions, we present the results only in two limits.

\vspace{12pt}
\noindent
(i) Long-wavelength limit ($k\eta/\sqrt{3} \ll 1$): In this case, we find
\begin{align}
\tau_{00} 
& = 
\frac{1}{8\pi G\eta^8} \left\{ - 39 |{d}_{2}|^{2}  
+ \mathcal{O} \left[ \left( \frac{k\eta}{\sqrt{3}} \right)^2 \right] \right\}
\, ,
\\
\tau_{ij} 
& = 
\frac{1}{8\pi G\eta^8} \delta_{ij} \left\{ - 19 |{d}_{2}|^{2}  
+ \mathcal{O} \left[ \left( \frac{k\eta}{\sqrt{3}} \right)^2 \right] \right\}
\, .
\end{align}
Recalling that $a(\eta) \propto \eta$ in RDE, for the most dominant terms, we have
\begin{align}
\varrho \propto \frac{1}{a^{10}}
\quad \text{and} \quad
\mathfrak{p} \propto \frac{1}{a^{10}}
\, ,
\end{align}
which decay quickly as the Universe expands.

\vspace{12pt}

\noindent
(ii) Short-wavelength limit ($k\eta/\sqrt{3} \gg 1$): In this case, we find
\begin{align}
\label{tau00LGshort} 
\tau_{00} 
& = 
\frac{1}{4\pi G \eta^{8}} \left\{ 3\Big( |{d}_{1}|^{2}  +  |{d}_{2}|^{2} 
\Big)  \left( \frac{k\eta}{\sqrt{3}} \right)^{6} 
+ \mathcal{O} \left[ \left( \frac{k\eta}{\sqrt{3}} \right)^5 \right] \right\}
\, ,
\\
\label{tauiiLGshort}
\tau_{ij} 
&= 
\frac{1}{4\pi G\eta^{8}} \delta_{ij}  \left\{ \Big( |{d}_{1}|^{2}  +  |{d}_{2}|^{2} 
\Big)  \left( \frac{k\eta}{\sqrt{3}} \right)^{6}  
+ \mathcal{O} \left[ \left( \frac{k\eta}{\sqrt{3}} \right)^5 \right] \right\}
\, .
\end{align}
We then have
\begin{align}
\label{eosLGshort}
\mathfrak{p}  \approx \frac{1}{3} \varrho
\propto \frac{1}{a^{4}}
\, ,
\end{align}
which behaves as radiation in this limit.

\subsection{Spatially-flat gauge}

Let us take the spatially-flat gauge by imposing the conditions
\begin{align}
\label{FG} 
\psi = E=0  \, .
\end{align}
From Eqs.~\eqref{Phi} and \eqref{Psi}, we have
\begin{align}
\Psi = \mathcal{H} Q \, .
\label{pfMflat}
\end{align}
As the gauge variables become $E=0$ and $Q=\Psi/{\cal H}$ in Eqs.~\eqref{gipf1}-\eqref{gipf3},
$\tau_{\mu\nu}$ is expressed only by the Bardeen variable $\Psi$ as
\begin{align}
\tau_{00} 
&= 
\frac{1}{8\pi G } \bigg[ 
-\frac{2}{\mathcal{H}^{\prime} -\mathcal{H}^{2} }\big\langle (\nabla \Psi^{\prime})^{2} \big\rangle 
- \frac{4\mathcal{H}}{\mathcal{H}^{\prime} -\mathcal{H}^{2} }\big\langle \nabla \Psi^{\prime}
\cdot \nabla \Psi \big\rangle 
+ \frac{(2\mathcal{H}^{\prime}-3\mathcal{H}^{2})(\mathcal{H}^{\prime} 
+\mathcal{H}^{2})} {\mathcal{H}^{2}(\mathcal{H}^{\prime} -\mathcal{H}^{2})} 
\big\langle (\nabla \Psi)^{2} \big\rangle            
\nonumber \\
& \hspace{3.5em} 
- 12 \big\langle (\Psi^{\prime})^{2} \big\rangle 
+ \frac{24(\mathcal{H}^{\prime} - 2\mathcal{H}^{2})}{\mathcal{H}} 
\big\langle \Psi^{\prime} \Psi \big\rangle 
-\frac{12(\mathcal{H}^{\prime} -2\mathcal{H}^{2})^{2}}
{\mathcal{H}^{2}}  \big\langle \Psi^{2} \big\rangle \bigg]   
\, ,    
\label{pfeffemfsM1}   
\\
\tau_{ij} 
&= 
\frac{1}{8\pi G  } \delta_{ij} \bigg[ \frac{2}{3\mathcal{H}^{2}}
\big\langle \nabla \Psi^{\prime \prime} \cdot \nabla \Psi \big\rangle  
+\frac{2(\mathcal{H}^{\prime} -2\mathcal{H}^{2})}{3\mathcal{H}^{2}(\mathcal{H}^{\prime} 
-\mathcal{H}^{2})}\big\langle (\nabla \Psi^{\prime})^{2} \big\rangle 
\nonumber \\
& \hspace{5em} 
- \frac{4(2\mathcal{H}^{\prime 2} -3\mathcal{H}^{\prime} \mathcal{H}^{2} +2\mathcal{H}^{4}  )}
{3\mathcal{H}^{3}(\mathcal{H}^{\prime} -\mathcal{H}^{2})}
\big\langle \nabla \Psi^{\prime} \cdot \nabla \Psi \big\rangle    
+\frac{8}{\mathcal{H}}  \big\langle \Psi^{\prime\prime} \Psi^{\prime} \big\rangle 
-\frac{8(\mathcal{H}^{\prime} -2\mathcal{H}^{2})}{\mathcal{H}^{2}}  
\big\langle \Psi^{\prime\prime} \Psi \big\rangle 
\nonumber \\
& \hspace{5em} 
-\frac{2\mathcal{H} (\mathcal{H}^{\prime} -\mathcal{H}^{2}) \mathcal{H}^{\prime \prime} 
- 6\mathcal{H}^{\prime 3} +10\mathcal{H}^{\prime 2} \mathcal{H}^{2} 
-3\mathcal{H}^{\prime} \mathcal{H}^{4} +\mathcal{H}^{6} } 
{3\mathcal{H}^{4} (\mathcal{H}^{\prime} -\mathcal{H}^{2}) }
\big\langle (\nabla \Psi)^{2} \big\rangle   
\nonumber \\
& \hspace{5em} 
-\frac{4(2\mathcal{H}^{\prime} -5\mathcal{H}^{2})}{\mathcal{H}^{2}}  
\big\langle (\Psi^{\prime} )^{2} \big\rangle    
-\frac{8( \mathcal{H} \mathcal{H}^{\prime \prime} 
- 2\mathcal{H}^{\prime 2} +3\mathcal{H}^{\prime} \mathcal{H}^{2} -6\mathcal{H}^{4} )} 
{\mathcal{H}^{3} }  \big\langle \Psi^{\prime} \Psi  \big\rangle 
\nonumber \\
& \hspace{5em} 
+ \frac{4(\mathcal{H}^{\prime} -2\mathcal{H}^{2})(2 \mathcal{H} \mathcal{H}^{\prime \prime}   
- 2\mathcal{H}^{\prime 2} -3\mathcal{H}^{\prime} \mathcal{H}^{2} -2\mathcal{H}^{4} ) }
{\mathcal{H}^{4}}  \big\langle \Psi^{2} \big\rangle     \bigg]     
\, .
\label{pfeffemfsM3}
\end{align}

\subsubsection{MDE}

Using the solution \eqref{eq:Psi-sol} for Eqs.~\eqref{pfeffemfsM1} and \eqref{pfeffemfsM3}, 
then we find
\begin{align}
\tau_{00} 
&= 
\frac{1}{8\pi G} \bigg[ \frac{4k^{2}}{3} |{c}_{1}|^{2}  
+ \mathcal{O} \left( \eta^{-2} \right) \bigg]
\, ,
\\
\tau_{ij}  
&= 
\frac{1}{8\pi G} \delta_{ij} \bigg[ \left(4  +\frac{17k^{2}}{18} \right) |{c}_{1}|^{2} 
+ \mathcal{O} \left( \eta^{-2} \right) \bigg]
\, .
\end{align}
For the most dominant term, we have the same  $a$-dependence \eqref{eosa2LG}
as in the longitudinal gauge
\begin{align}
\label{eosa2FG}
\varrho  \propto \frac{1}{a^2}
\quad \text{and} \quad
\mathfrak{p}  \propto \frac{1}{a^2}
\, ,
\end{align}
but with $\mathfrak{p} \approx (4+17k^2/18)/(4k^2/3) \varrho$.

\subsubsection{RDE}

Again, we consider two limits in RDE.

\vspace{12pt}
\noindent
(i) Long-wavelength limit ($k\eta/\sqrt{3} \ll 1$): We find in this limit
\begin{align}
\tau_{00} 
&= 
\frac{1}{8\pi G \eta^8} \left\{
9 |{d}_{2}|^{2} \left( \frac{k\eta}{\sqrt{3}} \right)^2 
+ \mathcal{O} \left[ \left( \frac{k\eta}{\sqrt{3}} \right)^4 \right] \right\}
\, ,
\\
\tau_{ij} 
&= 
\frac{1}{8\pi G \eta^6} \delta_{ij} \left\{ 4|{d}_{2}|^{2}
+ \mathcal{O} \left[ \left( \frac{k\eta}{\sqrt{3}} \right)^2 \right] \right\}
\, .
\end{align} 
Considering the most dominant term, the pressure $\mathfrak{p} \propto 1/a^8$ is more significant than
the energy density $\varrho \propto k^2/a^8$.

\vspace{12pt}

\noindent
(ii) Short-wavelength limit ($k\eta/\sqrt{3} \gg 1$): We obtain, in this limit, 
the most dominant terms are the same as Eqs.~\eqref{tau00LGshort} and 
\eqref{tauiiLGshort} in the longitudinal gauge,
and the equation of  state is also the same with Eq.~\eqref{eosLGshort},
which behaves as radiation.

\subsection{Comoving gauge}

Let us take the comoving gauge by imposing the conditions
\begin{align}
\beta_{,i} = v_{i} \quad \text{and} \quad E = 0 \, .
\end{align}
Using $v_{i}$ in Eq.~\eqref{gi1sthyd2}, we have
\begin{align}
Q =  \beta = \frac{\Psi^{\prime}+ \mathcal{H} \Psi }
{\mathcal{H}^{\prime} - \mathcal{H}^{2}} 
\, .
\end{align}
Similar to the previous gauge conditions, 
$\tau_{\mu\nu}$ is expressed only by the Bardeen variable $\Psi$ as
\begin{align}
\tau_{00} 
&= 
\frac{1}{8\pi G } \bigg[ 
-\frac{2\mathcal{H}}{(\mathcal{H}^{\prime} - \mathcal{H}^{2}  )^2} 
\big\langle \nabla \Psi^{\prime \prime} \cdot \nabla \Psi^{\prime}  \big\rangle
- \frac{2\mathcal{H}^{2}}{(\mathcal{H}^{\prime} -\mathcal{H}^{2})^{2} }
\big\langle \nabla \Psi^{\prime \prime} \cdot \nabla \Psi \big\rangle    
\nonumber \\
& 
+ \frac{2\mathcal{H} \big( \mathcal{H}^{\prime \prime} -3\mathcal{H} \mathcal{H}^{\prime}
+\mathcal{H}^{3} \big)}{(\mathcal{H}^{\prime} -\mathcal{H}^{2})^{3} }
\big\langle (\nabla \Psi^{\prime})^{2} \big\rangle      
+ \frac{4\mathcal{H} \big( \mathcal{H}\mathcal{H}^{\prime \prime} 
-2\mathcal{H}^{\prime 2} + \mathcal{H}^{2} \mathcal{H}^{\prime} - \mathcal{H}^{4} \big)}
{(\mathcal{H}^{\prime} -\mathcal{H}^{2})^{3}} 
\big\langle \nabla \Psi^{\prime } \cdot \nabla \Psi \big\rangle  
\nonumber \\
& 
+ \frac{5\mathcal{H}^{\prime 3} + 2\mathcal{H}^{3}\mathcal{H}^{\prime\prime}
- 23\mathcal{H}^{2}\mathcal{H}^{\prime 2} + 25\mathcal{H}^{4} \mathcal{H}^{\prime}
- 11\mathcal{H}^{6} } {(\mathcal{H}^{\prime} -\mathcal{H}^{2})^{3}} 
\big\langle (\nabla \Psi)^{2} \big\rangle       
- \frac{3 \mathcal{H}^{2}}{(\mathcal{H}^{\prime} -\mathcal{H}^{2})^{2}}
\big\langle \Psi^{\prime \prime 2} \big\rangle  
\nonumber \\
&  
 + \frac{6\mathcal{H}^{2} \big( \mathcal{H}^{\prime \prime} 
-2 \mathcal{H}\mathcal{H}^{\prime} \big)}{(\mathcal{H}^{\prime} -\mathcal{H}^{2})^{3}}    
\big\langle \Psi^{\prime \prime} \Psi^{\prime} \big\rangle  
+\frac{6 \mathcal{H}^{2} \big( \mathcal{H}\mathcal{H}^{\prime \prime}
- 4 \mathcal{H}^{\prime 2} +6\mathcal{H}^{2}\mathcal{H}^{\prime}
-4\mathcal{H}^{4} \big) }{(\mathcal{H}^{\prime} -\mathcal{H}^{2})^{3}}
\big\langle \Psi^{\prime \prime} \Psi \big\rangle      
\nonumber \\
&
- \frac{3 \mathcal{H}^{2} \big(\mathcal{H}^{\prime \prime}
- 4 \mathcal{H} \mathcal{H}^{\prime} +2 \mathcal{H}^{3} \big) 
\big(  \mathcal{H}^{\prime \prime} - 2 \mathcal{H}^{3} \big) }
{(\mathcal{H}^{\prime} -\mathcal{H}^{2})^{4}}
 \big\langle (\Psi^{\prime})^{2} \big\rangle    
\nonumber \\
& 
- \frac{ 6\mathcal{H}^{2} \big\{  \mathcal{H}\mathcal{H}^{\prime \prime 2}
-4\mathcal{H}^{\prime \prime } ( \mathcal{H}^{\prime 2} -\mathcal{H}^{2}\mathcal{H}^{\prime}
+\mathcal{H}^{4} ) + 12 \mathcal{H} \mathcal{H}^{\prime 3} 
-28 \mathcal{H}^{3}\mathcal{H}^{\prime 2} + 28 \mathcal{H}^{5} \mathcal{H}^{\prime}
- 8 \mathcal{H}^{7} \big\}  }{(\mathcal{H}^{\prime} -\mathcal{H}^{2})^{4}} 
\big\langle \Psi^{\prime} \Psi \big\rangle    
\nonumber \\
&
-\frac{ 3\mathcal{H}^{2} \big( \mathcal{H}\mathcal{H}^{\prime \prime}
-6\mathcal{H}^{\prime 2} +12\mathcal{H}^{2} \mathcal{H}^{\prime} -8\mathcal{H}^{4} \big) 
\big( \mathcal{H}\mathcal{H}^{\prime \prime } -2\mathcal{H}^{\prime 2} \big)  }
{(\mathcal{H}^{\prime} -\mathcal{H}^{2})^{4}} 
 \big\langle \Psi^{2} \big\rangle \bigg]   
\, ,    
\label{pfeffemcmM1}  
\end{align}
\begin{align}
\tau_{ij} 
&= 
\frac{1}{8\pi G  } \delta_{ij} \bigg[ 
\frac{2}{3(\mathcal{H}^{\prime} - \mathcal{H}^{2}  )^{2}} 
\big\langle \nabla \Psi^{\prime \prime \prime} \cdot \nabla \Psi^{\prime}  \big\rangle
+\frac{2\mathcal{H}}{3(\mathcal{H}^{\prime} - \mathcal{H}^{2}  )^{2}} 
\big\langle \nabla \Psi^{\prime \prime \prime} \cdot \nabla \Psi \big\rangle   
\nonumber \\
&
+ \frac{2}{3(\mathcal{H}^{\prime} -\mathcal{H}^{2})^{2} }
\big\langle (\nabla \Psi^{\prime \prime})^{2} \big\rangle      
-\frac{2  F_{1} }{3(\mathcal{H}^{\prime} - \mathcal{H}^{2}  )^{3}} 
\big\langle \nabla \Psi^{\prime \prime} \cdot \nabla \Psi^{\prime}  \big\rangle
-\frac{2 F_{2}} {3(\mathcal{H}^{\prime} - \mathcal{H}^{2}  )^{3}} 
\big\langle \nabla \Psi^{\prime \prime} \cdot \nabla \Psi \big\rangle   
\nonumber \\
&
- \frac{  2 F_{3} }{3(\mathcal{H}^{\prime} -\mathcal{H}^{2})^{4} } 
\big\langle (\nabla \Psi^{\prime})^{2} \big\rangle      
- \frac{2 F_{4}}{3(\mathcal{H}^{\prime} -\mathcal{H}^{2})^{4}} 
\big\langle \nabla \Psi^{\prime } \cdot \nabla \Psi \big\rangle   
- \frac{F_{5}} {3(\mathcal{H}^{\prime} -\mathcal{H}^{2})^{4}} 
\big\langle (\nabla \Psi)^{2} \big\rangle             
\nonumber \\
&
+ \frac{2 \mathcal{H}}{(\mathcal{H}^{\prime} -\mathcal{H}^{2})^{2}}
\big\langle \Psi^{\prime \prime \prime} \Psi^{\prime \prime} \big\rangle
- \frac{2F_{6}}{(\mathcal{H}^{\prime} -\mathcal{H}^{2})^{3}}    
\big\langle \Psi^{\prime \prime\prime} \Psi^{\prime} \big\rangle  
- \frac{2 F_{7} }{(\mathcal{H}^{\prime} -\mathcal{H}^{2})^{3}}
\big\langle \Psi^{\prime \prime \prime} \Psi \big\rangle 
- \frac{F_{8} }{(\mathcal{H}^{\prime} -\mathcal{H}^{2})^{3}}
\big\langle \Psi^{\prime \prime 2} \big\rangle 
\nonumber \\
&
+ \frac{2 F_{9} }{(\mathcal{H}^{\prime} -\mathcal{H}^{2})^{4}}
\big\langle \Psi^{\prime \prime } \Psi^{\prime} \big\rangle      
- \frac{2 F_{10} }{(\mathcal{H}^{\prime} -\mathcal{H}^{2})^{4}}
\big\langle \Psi^{\prime \prime } \Psi \big\rangle 
+\frac{F_{11}}{(\mathcal{H}^{\prime} -\mathcal{H}^{2})^{5}}
 \big\langle (\Psi^{\prime})^{2} \big\rangle     
- \frac{ 2F_{12}  }{(\mathcal{H}^{\prime} -\mathcal{H}^{2})^{5}} 
\big\langle \Psi^{\prime} \Psi \big\rangle  
\nonumber \\
&
+\frac{ F_{13}  }{(\mathcal{H}^{\prime} -\mathcal{H}^{2})^{5}} 
\big\langle \Psi^{2}  \big\rangle      \bigg]    
\, ,
\label{pfeffemcmM3}
\end{align} 
where
\begin{align}
F_{1} 
& =  
4\mathcal{H}^{\prime \prime} -13\mathcal{H} \mathcal{H}^{\prime} + 5\mathcal{H}^{3}  
\, , 
\\
F_{2} 
& =  4\mathcal{H} \mathcal{H}^{\prime \prime} -2 \mathcal{H}^{\prime 2} 
- 9 \mathcal{H}^{2} \mathcal{H}^{\prime}  +3\mathcal{H}^{4}   
\, , 
\\
F_{3} 
& = 
(\mathcal{H}^{\prime}-\mathcal{H}^{2}) \mathcal{H}^{\prime \prime \prime} 
- 3 \mathcal{H}^{\prime \prime 2}- 3 \mathcal{H}^{\prime 3}   + 4 \mathcal{H} 
(4\mathcal{H}^{\prime} -\mathcal{H}^{2}  ) \mathcal{H}^{\prime \prime} 
-24 \mathcal{H}^{2} \mathcal{H}^{\prime 2} + 19\mathcal{H}^{4} \mathcal{H}^{\prime}
-4\mathcal{H}^{6} 
\, , 
\\
F_{4} 
& = 2 \mathcal{H}(\mathcal{H}^{\prime} - \mathcal{H}^{2}) 
\mathcal{H}^{\prime \prime \prime} - ( 6\mathcal{H} \mathcal{H}^{\prime \prime} 
- 3\mathcal{H}^{\prime 2} -26\mathcal{H}^{2}\mathcal{H}^{\prime} 
+ 5\mathcal{H}^{4}  ) \mathcal{H}^{\prime \prime} 
- 4\mathcal{H} ( 4\mathcal{H}^{\prime 3} +6\mathcal{H}^{2}\mathcal{H}^{\prime 2} 
\nonumber \\
& \hspace{1em}
-5\mathcal{H}^{4} \mathcal{H}^{\prime} +\mathcal{H}^{6} ) 
\, ,  
\\
F_{5} 
& = 
2 \mathcal{H}^{2} (\mathcal{H}^{\prime} - \mathcal{H}^{2} ) 
\mathcal{H}^{\prime \prime \prime}
+ 2 \mathcal{H} ( 3\mathcal{H}^{\prime 2} + 10\mathcal{H}^{2} \mathcal{H}^{\prime}
-\mathcal{H}^{4}  ) \mathcal{H}^{\prime \prime} 
- 6\mathcal{H}^{2} \mathcal{H}^{\prime \prime 2} -\mathcal{H}^{\prime 4} 
- 22\mathcal{H}^{2} \mathcal{H}^{\prime 3}  
\nonumber \\
& \hspace{1em}
 - 6 \mathcal{H}^{4} \mathcal{H}^{\prime 2} +6 \mathcal{H}^{6} \mathcal{H}^{\prime}
- \mathcal{H}^{8} 
\, ,   
\\
F_{6} 
& = 
\mathcal{H} (\mathcal{H}^{\prime \prime} 
- 2 \mathcal{H} \mathcal{H}^{\prime} ) 
\, ,   
\\
F_{7} 
& = 
\mathcal{H} (\mathcal{H} \mathcal{H}^{\prime \prime} 
-4 \mathcal{H}^{\prime 2} + 6\mathcal{H}^{2} \mathcal{H}^{\prime} 
- 4 \mathcal{H}^{4}  ) 
\, ,  
\\
F_{8} 
& = 
4 \mathcal{H} \mathcal{H}^{\prime \prime} - 4\mathcal{H}^{\prime 2}
-3 \mathcal{H}^{2} \mathcal{H}^{\prime} - \mathcal{H}^{4} 
\, ,  
\\
F_{9} 
& = 
4 \mathcal{H}\mathcal{H}^{\prime \prime 2} 
- \mathcal{H} ( \mathcal{H}^{\prime} -\mathcal{H}^{2} ) 
\mathcal{H}^{\prime \prime \prime} + 20 \mathcal{H} \mathcal{H}^{\prime 3}
-2 (2 \mathcal{H}^{\prime 2} + 5\mathcal{H}^{2} \mathcal{H}^{\prime}
+ \mathcal{H}^{4}  ) \mathcal{H}^{\prime \prime} 
-28 \mathcal{H}^{3} \mathcal{H}^{\prime 2}   
\nonumber \\
& \hspace{1em}
+ 38 \mathcal{H}^{5} \mathcal{H}^{\prime} - 14 \mathcal{H}^{7} 
\, , 
\\
F_{10} 
& = 
\mathcal{H}^{2} ( \mathcal{H}^{\prime} -\mathcal{H}^{2} ) 
\mathcal{H}^{\prime \prime \prime} 
- 4 \mathcal{H}^{2} \mathcal{H}^{\prime \prime 2}
+ \mathcal{H} ( 13 \mathcal{H}^{\prime 2} -8 \mathcal{H}^{2}
\mathcal{H}^{\prime} +11\mathcal{H}^{4}  ) \mathcal{H}^{\prime \prime}
-10 \mathcal{H}^{\prime 4}
-10 \mathcal{H}^{2} \mathcal{H}^{\prime 3}         
\nonumber \\
& \hspace{1em}
- 52 \mathcal{H}^{6} \mathcal{H}^{\prime}  
+ 40 \mathcal{H}^{4} \mathcal{H}^{\prime 2} + 16 \mathcal{H}^{8} 
\, , 
\\
F_{11} 
& = 2 \mathcal{H} ( \mathcal{H}^{\prime} -\mathcal{H}^{2} )
( \mathcal{H}^{\prime \prime} -2\mathcal{H}\mathcal{H}^{\prime} )
\mathcal{H}^{\prime \prime \prime} 
- 4 \mathcal{H} \mathcal{H}^{\prime \prime 3}
+ ( 4\mathcal{H}^{\prime 2} + 17 \mathcal{H}^{2} \mathcal{H}^{\prime}
+ 3 \mathcal{H}^{4} ) \mathcal{H}^{\prime \prime 2}         
\nonumber \\
&  \quad
- 4\mathcal{H} ( 10\mathcal{H}^{\prime 3} - 11 \mathcal{H}^{2}
\mathcal{H}^{\prime 2}+20\mathcal{H}^{4} \mathcal{H}^{\prime}
-7 \mathcal{H}^{6} ) \mathcal{H}^{\prime \prime} + 4\mathcal{H}^{2}
( 18 \mathcal{H}^{\prime 4} -46 \mathcal{H}^{2} \mathcal{H}^{\prime 3}
+70 \mathcal{H}^{4} \mathcal{H}^{\prime 2}        
\nonumber \\
& \hspace{1em}
-43 \mathcal{H}^{6} \mathcal{H}^{\prime} + 9\mathcal{H}^{8} ) 
\, , 
\\
F_{12} 
& = 
4 \mathcal{H}^{2} \mathcal{H}^{\prime \prime 3} 
- 2\mathcal{H}(\mathcal{H}^{\prime} -\mathcal{H}^{2}  )
( \mathcal{H}\mathcal{H}^{\prime \prime} -2\mathcal{H}^{\prime 2}
+2\mathcal{H}^{2}\mathcal{H}^{\prime} -2\mathcal{H}^{4}  )
\mathcal{H}^{\prime \prime \prime}         
\nonumber \\
& \quad
 -\mathcal{H} ( 13\mathcal{H}^{\prime 2} 
-\mathcal{H}^{2} \mathcal{H}^{\prime} + 12 \mathcal{H}^{4} )
 \mathcal{H}^{\prime \prime 2} 
 + 2 ( 5\mathcal{H}^{\prime 4} +22\mathcal{H}^{2} \mathcal{H}^{\prime 3}
 -40 \mathcal{H}^{4} \mathcal{H}^{\prime 2} 
 + 50 \mathcal{H}^{6} \mathcal{H}^{\prime} 
 \nonumber \\
& \hspace{1em}
 -  13\mathcal{H}^{8} ) \mathcal{H}^{\prime \prime}         
- 4 \mathcal{H} ( 15 \mathcal{H}^{\prime 5} -28 \mathcal{H}^{2} \mathcal{H}^{\prime 4}
+ 16 \mathcal{H}^{4} \mathcal{H}^{\prime 3} + 19 \mathcal{H}^{6} \mathcal{H}^{\prime 2}
-18 \mathcal{H}^{8} \mathcal{H}^{\prime} + 4\mathcal{H}^{10} ) 
\, ,  
\\
F_{13} 
& = 
2\mathcal{H}^{2} ( \mathcal{H}^{\prime} -\mathcal{H}^{2}) 
( \mathcal{H} \mathcal{H}^{\prime \prime} - 4\mathcal{H}^{\prime 2}
+ 6\mathcal{H}^{2} \mathcal{H}^{\prime} - 4\mathcal{H}^{4} ) 
\mathcal{H}^{\prime \prime \prime} -4 \mathcal{H}^{3} \mathcal{H}^{\prime \prime 3}
+ (22 \mathcal{H}^{2} \mathcal{H}^{\prime 2} -19 \mathcal{H}^{4} \mathcal{H}^{\prime}
\nonumber \\
& \quad
 +21 \mathcal{H}^{6}  ) \mathcal{H}^{\prime \prime 2}  
-4 \mathcal{H} ( 7\mathcal{H}^{\prime 4} + 4\mathcal{H}^{2} \mathcal{H}^{\prime 3}
-17 \mathcal{H}^{4} \mathcal{H}^{\prime 2} + 22 \mathcal{H}^{6} \mathcal{H}^{\prime}
-4 \mathcal{H}^{8} ) \mathcal{H}^{\prime \prime}      
\nonumber \\
& \quad
 + 4 ( 6\mathcal{H}^{\prime 4} +3\mathcal{H}^{2} \mathcal{H}^{\prime 3} 
-25 \mathcal{H}^{4} \mathcal{H}^{\prime 2} + 32 \mathcal{H}^{6} \mathcal{H}^{\prime}
- 8\mathcal{H}^{8} ) \mathcal{H}^{\prime 2} 
\, .
\end{align}

\subsubsection{MDE}

Plugging the solution \eqref{eq:Psi-sol} in Eqs.~\eqref{pfeffemcmM1} and \eqref{pfeffemcmM3}, 
then we obtain
\begin{align}
\tau_{00} 
&=  
\frac{1}{8\pi G} \bigg[ \frac{77k^{2}}{9} |{c}_{1}|^{2} 
+ \mathcal{O}(\eta^{-5}) \bigg]
\, ,
\\
\tau_{ij} 
&=  \frac{1}{8\pi G} \delta_{ij} 
\bigg[ \left(\frac{16}{9} +\frac{13k^{2}}{27} \right)|{c}_{1}|^{2} 
+ \mathcal{O}(\eta^{-2}) \bigg]
\, ,
\end{align}
which give the same $a$-dependence 
as in the other gauges, Eqs.~\eqref{eosa2LG} and \eqref{eosa2FG}, 
but the equation of state becomes
$\mathfrak{p} \approx (16+13k^2/3)/(77k^2) \varrho$.

\subsubsection{RDE}

Again, we consider two limits in RDE.

\vspace{12pt}

\noindent
(i) Long-wavelength limit ($k\eta/\sqrt{3} \ll 1$): The most dominant terms are given by
\begin{align}
\tau_{00} 
& =  
\frac{1}{8\pi G \eta^{8}} \left\{
\frac{45}{2} |{d}_{2}|^{2}  
\left( \frac{k\eta}{\sqrt{3}} \right)^2 
+ \mathcal{O} \left[ \left( \frac{k\eta}{\sqrt{3}} \right)^4 \right] \right\}
\, ,
\\
\tau_{ij} 
& = 
\frac{1}{8\pi G \eta^6} \delta_{ij} \left\{ |{d}_{2}|^{2}
+ \mathcal{O} \left[ \left( \frac{k\eta}{\sqrt{3}} \right)^2 \right] \right\}
\, .
\end{align}
Similar to the spatially-flat gauge case,
the pressure $\mathfrak{p} \propto 1/a^8$ is more significant than
the energy density $\varrho \propto k^2/a^8$.

\vspace{12pt}

\noindent
(ii) Short-wavelength limit ($k\eta/\sqrt{3} \gg 1$): The dominant terms are given by
\begin{align}
\tau_{00} 
&= 
\frac{1}{4\pi G\eta^{8}}  \left\{ \frac{45}{2}
\Big(  |{d}_{1}|^{2} +  |{d}_{2}|^{2}  \Big)  \left( \frac{k\eta}{\sqrt{3}} \right)^{4} 
+ \mathcal{O} \left[ \left( \frac{k\eta}{\sqrt{3}} \right)^3 \right] \right\}
\, ,
\\
\tau_{ij} 
&= 
\frac{1}{4\pi G \eta^8} \delta_{ij}  \left\{ \frac{3}{2} \Big(  |{d}_{1}|^{2}+  |{d}_{2}|^{2}  \Big)
 \left( \frac{k\eta}{\sqrt{3}} \right)^{4} 
+ \mathcal{O} \left[ \left( \frac{k\eta}{\sqrt{3}} \right)^3 \right] \right\}
\, .
\end{align}
For the most dominant terms, $\varrho \propto 1/a^6$ and $\mathfrak{p} \propto 1/a^6$ 
and the equation of state is $\mathfrak{p} \approx \varrho/15$.

\section{Conclusions}
\label{secconc}

In this work, we have investigated the gauge invariance of 2EMT in the Friedmann universe.
Introducing the gravitational scalar perturbations as well as the matter ones of fluid, we have 
kept the contributions up to second order in perturbations.
The second-order terms consist of two parts:
(i) the linear terms in second-order perturbations, and 
(ii) the quadratic combinations of first-order perturbations.
In the Einstein tensor, 
(i) $G^{(1)}_{\mu\nu}[g^{(2)}]$ is regarded as a pure second-order geometric contribution, and 
(ii) $G^{(2)}_{\mu\nu}[g^{(1)}]$ is regarded as a contribution to the effective energy-momentum 
tensor along with $T^{(2)}_{\mu\nu}[g^{(1)},\delta\rho^{(1)},\delta p^{(1)}]$.  
As a result, 2EMT is given by
$T^{(2,\rm{eff})}_{\mu\nu} = T^{(2)}_{\mu\nu}[g^{(1)},\delta\rho^{(1)},\delta p^{(1)}] 
- G^{(2)}_{\mu\nu}[g^{(1)}]/8\pi G$.
Finally we have integrated over several wavelengths to obtain
$\tau_{\mu\nu} = \left\langle T^{(2,\rm{eff})}_{\mu\nu} \right\rangle$.

Using the first-order equations, we could evaluate $\tau_{\mu\nu}$ in terms of the gravitational 
perturbations in Section~\ref{sectau}. The result shows that $\tau_{\mu\nu}$ depends not only 
on the gauge-invariant Bardeen variable $\Psi$, but also the gauge variables, $Q=\beta+E$ and $E$.
This indicates that $\tau_{\mu\nu}$ is definitely gauge dependent. The fact that $\tau_{\mu\nu}$ 
is not gauge invariant is not unreasonable: In general, the tensor components change after gauge 
(or infinitesimal coordinate) transformations. Even the rank-zero tensor (scalar) changes if it is 
a local quantity, e.g. the energy density. Therefore, the gauge dependence of 2EMT of 
cosmological scalar perturbations is in fact not exceptional.

Any truly observable quantity is supposed to be gauge invariant from the beginning. In our case, 
$\tau_{\mu\nu}$ is not in this category. However, if the value of the quantity converges in a certain 
limit after imposing gauge conditions, we may hope that it may have some (in)direct connection to 
observables. In this sense, we have examined $\tau_{\mu\nu}$ in three gauge choices --
longitudinal, spatially-flat, and comoving gauges.

Once we select a gauge condition, $\tau_{\mu\nu}$ can be expressed only by the gauge invariant 
variable $\Psi$. However, the functional form of $\tau_{\mu\nu}$ is dependent on the gauge choice.
In order to investigate $\tau_{\mu\nu}$ in certain limits, we have performed the Fourier-mode 
expansion for $\Psi$, and have solved the equation for $\Psi$ in the matter- and radiation-dominated 
epoch. Plugging the solution for $\Psi$ in $\tau_{\mu\nu}$, the results show that $\tau_{\mu\nu}$ 
never converges in any wavelength limit. In addition, interpreting the components of $\tau_{\mu\nu}$
as the effective second-order energy density and pressure, 
$\tau^\mu{}_\nu = \text{diag}(-\varrho,\mathfrak{p},\mathfrak{p},\mathfrak{p})$,
the effective equation-of-state parameter  $\mathfrak{p}/\varrho$ does not converge to a single value 
in any limit of three gauge choices. As a conclusion, $\tau_{\mu\nu}$ and its effects are strictly gauge
dependent.

We may also well quantize $\Psi$ as
\begin{align}
\Psi(\eta,{\bm x}) 
=  \int \frac{d^3k}{(2\pi)^{3/2}} 
\left[ \hat{\Psi}_ {\bm{k}}(\eta) e^{i \bm{k} \cdot \bm{x}} \hat{a}_ {\bm{k}}
+ \hat{\Psi}^*_ {\bm{k}}(\eta) e^{-i \bm{k} \cdot \bm{x}} \hat{a}^{\dag}_ {\bm{k}} \right]
\, ,
\end{align}
where the annihilation and the creation operators satisfy the usual commutator relations
with proper normalizations. 
Solving the field equation \eqref{Psieq} for $\Psi$ with this quantization,
we find the same solutions for $\hat{\Psi}_ {\bm{k}}(\eta)$ as Eq. \eqref{eq:Psi-sol}.
%as in the case of Fourier-mode expansion .
Using the ordering $\Psi^{(m)}\Psi^{(n)} = \big(\Psi^{(m)}\Psi^{(n)} + \Psi^{(n)}\Psi^{(m)}\big)/2$,
and applying the operator on the ground state,
we find the same $\tau_{\mu\nu}$ as obtained in Section~\ref{sec3}.
Therefore, $\tau_{\mu\nu}$ has no quantum effect, 
or at least any quantum contributions are indistinguishable from classical ones
(see also Refs. \cite{Maggiore:2010wr,Maggiore:2011hw,Hollenstein:2011cz,
Glavan:2013mra,Glavan:2014uga,Glavan:2015cut,Aoki:2014ita,Aoki:2014dqa}
for the quantum effects of the backreaction).
%The reason is that we are considering the quadratic order in the energy-momentum tensor,
%and the quantum effect usually arises in the higher orders. 
%(There have been some works for investigating the quantum effects of the back-reaction
%for several fields in the Universe, for example, 
%Refs.~\cite{Maggiore:2010wr,Maggiore:2011hw,Hollenstein:2011cz,
%Glavan:2013mra,Glavan:2014uga,Glavan:2015cut,Aoki:2014ita,Aoki:2014dqa}.)

\acknowledgments

We are grateful to Gungwon Kang, Jai-chan Hwang, Viatcheslav Mukhanov and Alexei Starobinsky
for helpful discussions.
This work was supported by the grant from the National Research Foundation
funded by the Korean government, NRF-2017R1A2B4010738 (I.C.),
NRF-2019R1A2C2085023 (J.G.),
and NRF-2018-R1D1A1B0-7048945 (S.H.O.).
J.G. also acknowledges the Korea-Japan Basic Scientific Cooperation Program 
supported by the National Research Foundation of Korea 
and the Japan Society for the Promotion of Science (2018K2A9A2A08000127),
and the Asia Pacific Center for Theoretical Physics 
for hospitality while this work was completed
and for Focus Research Program ``The origin and evolution of the Universe'' 
where parts of this work were presented and discussed.

\appendix

\section{2EMT during MDE and RDE}
\label{app:2emt-mdrd}

In this section, we present more complete functional forms of 2EMT during MDE and RDE in the three gauge conditions presented in the main text. In this section, for notational simplicity we define
$k\eta / \sqrt{3} \equiv \mathcal{T}$, $\sin(k\eta / \sqrt{3}) \equiv \mathcal{S}$, 
$\cos(k\eta / \sqrt{3}) \equiv \mathcal{C}$ and 
$\big({c}_{1},{c}_{2}\big) \equiv {c}_{1}^*{c}_{2} + {c}_{1}{c}_{2}^* = 2\Re\big({c}_{1}^*{c}_{2}\big)$.

\subsection{Longitudinal gauge}

\subsubsection{MDE}

\begin{align}
\tau_{00} 
= & 
\frac{1}{8\pi G} \bigg[ \frac{19k^{2}}{3} |{c}_{1}|^{2}  
- \frac{48}{\eta^{2}} |{c}_{1}|^{2}  + \frac{3k^{2}}{\eta^{5}} \big({c}_{1},{c}_{2}\big)  
- \frac{48}{\eta^{7}} \big({c}_{1},{c}_{2}\big) +\frac{8k^{2}}{\eta^{10}} |{c}_{2}|^{2} 
-\frac{123}{\eta^{12}} |{c}_{2}|^{2}   \bigg]    
\, , 
\label{dustlg00}  
\\
\tau_{ij} 
= & 
\frac{1}{8\pi G} \delta_{ij} \bigg[ \frac{k^{2}}{9} |{c}_{1}|^{2}  
- \frac{k^{2}}{\eta^{5}} \big({c}_{1},{c}_{2}\big)  
- \frac{40}{\eta^{7}} \big({c}_{1},{c}_{2}\big) 
+\frac{2k^{2}}{3\eta^{10}} |{c}_{2}|^{2} 
-\frac{55}{\eta^{12}} |{c}_{2}|^{2}   \bigg]    
\, .
\label{dustlgij} 
\end{align}

\subsubsection{RDE in long-wavelength limit}

 \begin{align}
\tau_{00} &=  \frac{1}{8\pi G\eta^8} \Big[ - 39 |d_{2}|^{2}  
+ 6 |d_{2}|^{2}  \mathcal{T}^{2}
+ 4 \big(d_{1},d_{2}\big)   \mathcal{T}^{3}
+12|d_{2}|^{2}  \mathcal{T}^{4}   
- \frac{4 \big(d_{1},d_{2}\big) }{5}  \mathcal{T}^{5}
\notag \\
& \hspace{.4in}
+ \frac{13|d_{2}|^{2} - 4|d_{1}|^{2}}{3}  \mathcal{T}^{6}
- \frac{113 \big(d_{1},d_{2}\big) }{35}  \mathcal{T}^{7}  
+  \frac{34\big(|d_{1}|^{2} - |d_{2}|^{2} \big) }{15} \mathcal{T}^{8}
+ {\cal O} \Big( \mathcal{T}^{9} \Big)  \Big]    , \label{lwlg00}  \\
\tau_{ij} &= \frac{1}{8\pi G\eta^8} \delta_{ij} \Big[ - 19 |d_{2}|^{2}  
- 14 |d_{2}|^{2}  \mathcal{T}^{2}
+ \frac{16 \big(d_{1},d_{2}\big) }{3}  \mathcal{T}^{3}  
+ \frac{28 \big(d_{1},d_{2}\big) }{15}  \mathcal{T}^{5} 
\notag \\
& \hspace{.4in}
+ \frac{13|d_{2}|^{2} - 4|d_{1}|^{2}}{9}  \mathcal{T}^{6}
- \frac{3 \big(d_{1},d_{2}\big) }{35}  \mathcal{T}^{7}   
-  \frac{4\big(|d_{1}|^{2} - |d_{2}|^{2} \big) }{45} \mathcal{T}^{8} 
+ {\cal O} \Big( \mathcal{T}^{9} \Big)    \Big]    . \label{lwlgij} 
\end{align}

\subsubsection{RDE in short-wavelength limit}

\begin{align}
\tau_{00} =  \frac{1}{8\pi G \eta^{8}} \Big[ &  3\Big(\mathcal{S}^{2} |d_{1}|^{2}  + \mathcal{C}^{2}  |d_{2}|^{2} 
-\mathcal{SC}  \big(d_{1},d_{2}\big)   \Big)  \mathcal{T}^{6}   \notag \\
& + \Big(12 \mathcal{SC}\big( |d_{1}|^{2}  -  |d_{2}|^{2} \big) - 6 \big( \mathcal{C}^{2} - \mathcal{S}^{2}\big)  
  \big(d_{1},d_{2}\big)   \Big)  \mathcal{T}^{5} \notag \\ 
& + \Big(\big(27 \mathcal{C}^{2} -15 \mathcal{S}^{2} \big) |d_{1}|^{2} + \big(27 \mathcal{S}^{2} -15 \mathcal{C}^{2} \big) |d_{2}|^{2} 
+42 \mathcal{SC}  \big(d_{1},d_{2}\big)    \Big)  \mathcal{T}^{4}     \notag \\ 
& + \Big(72 \mathcal{SC} \big( |d_{2}|^{2} - |d_{1}|^{2}\big)  
+36 \big(\mathcal{C}^{2} -\mathcal{S}^{2} \big) \big(d_{1},d_{2}\big)   \Big)  \mathcal{T}^{3}  \notag \\ 
& + \Big(-84 \mathcal{SC}  \big(d_{1},d_{2}\big) - 39 \big(\mathcal{C}^{2} |d_{1}|^{2} + \mathcal{S}^{2} |d_{2}|^{2} \big)
+45 \big(\mathcal{S}^{2} |d_{1}|^{2} + \mathcal{C}^{2} |d_{2}|^{2} \big)  \Big)  \mathcal{T}^{2}  \notag \\ 
& + \Big( - 78 \mathcal{SC} \big( |d_{1}|^{2} - |d_{2}|^{2}\big)  
+ 39 \big(\mathcal{S}^{2}-\mathcal{C}^{2}\big) \big(d_{1},d_{2}\big)   \Big)  \mathcal{T}    \notag \\ 
& + \Big( - 39 \big(\mathcal{S}^{2} |d_{1}|^{2} + \mathcal{C}^{2} |d_{2}|^{2} \big)
+ 39 \mathcal{SC} \big(d_{1},d_{2}\big)   \Big)  \Big]    , \label{swlg00}  \\
\tau_{ij} = \frac{1}{8\pi G\eta^{8}} \delta_{ij}  \Big[ &  \Big(\mathcal{S}^{2} |d_{1}|^{2}  + \mathcal{C}^{2}  |d_{2}|^{2} 
-\mathcal{SC}  \big(d_{1},d_{2}\big)   \Big)  \mathcal{T}^{6}   \notag \\
& + \Big(4 \mathcal{SC}\big( |d_{1}|^{2}  -  |d_{2}|^{2} \big) - 2 \big( \mathcal{C}^{2} - \mathcal{S}^{2}\big)  
  \big(d_{1},d_{2}\big)   \Big)  \mathcal{T}^{5} \notag \\ 
& + \Big(3 \big(\mathcal{C}^{2}-\mathcal{S}^{2}\big) \big( |d_{1}|^{2}  -  |d_{2}|^{2} \big) 
+6 \mathcal{SC}  \big(d_{1},d_{2}\big)    \Big)  \mathcal{T}^{4}     \notag \\ 
& + \Big(8 \mathcal{SC} \big( |d_{2}|^{2} - |d_{1}|^{2}\big)  
+4 \big(\mathcal{C}^{2} -\mathcal{S}^{2} \big) \big(d_{1},d_{2}\big)   \Big)  \mathcal{T}^{3}  \notag \\ 
& + \Big(-24 \mathcal{SC}  \big(d_{1},d_{2}\big) - 19 \big(\mathcal{C}^{2} |d_{1}|^{2} + \mathcal{S}^{2} |d_{2}|^{2} \big)
+5 \big(\mathcal{S}^{2} |d_{1}|^{2} + \mathcal{C}^{2} |d_{2}|^{2} \big)  \Big)  \mathcal{T}^{2}  \notag \\ 
& + \Big( 38 \mathcal{SC} \big( |d_{1}|^{2} - |d_{2}|^{2}\big)  
+ 19 \big(\mathcal{S}^{2}-\mathcal{C}^{2}\big) \big(d_{1},d_{2}\big)   \Big)  \mathcal{T}    \notag \\ 
& + \Big( - 19 \big(\mathcal{S}^{2} |d_{1}|^{2} + \mathcal{C}^{2} |d_{2}|^{2} \big)
+ 19 \mathcal{SC} \big(d_{1},d_{2}\big)   \Big)  \Big]   .  \label{swlgij} 
\end{align}

\subsection{Spatially-flat gauge}

\subsubsection{MDE}

\begin{align}
\tau_{00} 
&=  
\frac{1}{8\pi G} \bigg[ \frac{4k^{2}}{3} |{c}_{1}|^{2}  
- \frac{300}{\eta^{2}} |{c}_{1}|^{2}  - \frac{2k^{2}}{\eta^{5}} \big({c}_{1},{c}_{2}\big)  
+\frac{3k^{2}}{\eta^{10}} |{c}_{2}|^{2}  \bigg]   
\, , 
\label{dustfg00}  
\\
\tau_{ij} 
&=  
\frac{1}{8\pi G} \delta_{ij} \bigg[ 4 |{c}_{1}|^{2}  
+\frac{17k^{2}}{18} |{c}_{1}|^{2}  - \frac{16}{\eta^{2}} |{c}_{1}|^{2}  
+ \frac{4- k^{2}}{\eta^{5}} \big({c}_{1}, {c}_{2}\big)  
- \frac{16}{\eta^{7}} \big({c}_{1},{c}_{2}\big) 
+\frac{4(k^{2}+1)}{\eta^{10}} |{c}_{2}|^{2}   
\nonumber \\
& \hspace{5em}
-\frac{16}{\eta^{12}} |{c}_{2}|^{2}   \bigg]   
\, . 
\label{dustfgij} 
\end{align}

\subsubsection{RDE in long-wavelength limit}

\begin{align}
\tau_{00} = \frac{1}{8\pi G} \Big[ &
\frac{9}{\eta^{8}} |d_{2}|^{2}  \mathcal{T}^{2}
-\frac{15}{\eta^{8}} |d_{2}|^{2}  \mathcal{T}^{4}
+ \frac{15 \big(d_{1},d_{2}\big) }{\eta^{8}}  \mathcal{T}^{5} 
+ \frac{15|d_{2}|^{2} - 12|d_{1}|^{2}}{\eta^{8}}  \mathcal{T}^{6}
 \nonumber\\
&
- \frac{41 \big(d_{1},d_{2}\big) }{5\eta^{8}}  \mathcal{T}^{7} 
+  \frac{4\big(|d_{1}|^{2} - |d_{2}|^{2} \big) }{\eta^{8}} \mathcal{T}^{8}   
+ {\cal O} \Big( \mathcal{T}^{9} \Big)  \Big]    , \label{lwfg00}  \\
\tau_{ij} = \frac{1}{8\pi G}  \delta_{ij} \Big[ & 4 \Big( \frac{1}{\eta^{6}}- \frac{1}{\eta^{8}} \Big) |d_{2}|^{2}
+\Big( \frac{4}{\eta^{6}}+ \frac{11}{\eta^{8}} \Big)  |d_{2}|^{2}  \mathcal{T}^{2}
-\frac{4}{3}\Big( \frac{1}{\eta^{6}}- \frac{1}{\eta^{8}} \Big)  \big(d_{1},d_{2}\big) 
 \mathcal{T}^{3}           
 \nonumber \\
&
- \frac{11}{\eta^{8}}|d_{2}|^{2} \mathcal{T}^{4}  
- \frac{1}{15}\Big( \frac{8}{\eta^{6}}- \frac{143}{\eta^{8}} \Big)  \big(d_{1},d_{2}\big) 
 \mathcal{T}^{5}        
+ \Big( \frac{4}{9\eta^{6}}\big( |d_{1}|^{2} - |d_{2}|^{2} \big) 
\nonumber \\
&
 - \frac{1}{9\eta^{8}} \big( 40|d_{1}|^{2} - 49|d_{2}|^{2} \big)    \Big)  
 \mathcal{T}^{6}       
 + \frac{1}{35} \Big( \frac{8}{\eta^{6}} - \frac{59}{3\eta^{8}} \Big)   \big(d_{1},d_{2}\big) 
 \mathcal{T}^{7}
 \nonumber \\ 
&
 -\frac{1}{45} \Big( \frac{4}{\eta^{6}} + \frac{26}{\eta^{8}} \Big) \big( |d_{1}|^{2} - |d_{2}|^{2} \big) 
  \mathcal{T}^{8}       
+ {\cal O} \Big( \mathcal{T}^{9} \Big)    \Big]    . \label{lwfgij} 
\end{align}

\subsubsection{RDE in short-wavelength limit}

\begin{align}
\tau_{00} = \frac{1}{8\pi G \eta^{8}} \Big[& 3\Big(\mathcal{S}^{2} |d_{1}|^{2}  + \mathcal{C}^{2}  |d_{2}|^{2} 
-\mathcal{SC}  \big(d_{1},d_{2}\big)   \Big)  \mathcal{T}^{6}   \notag \\
&+ \Big(12 \mathcal{SC}\big( |d_{1}|^{2}  -  |d_{2}|^{2} \big) - 6 \big( \mathcal{C}^{2} - \mathcal{S}^{2}\big)  
  \big(d_{1},d_{2}\big)   \Big)  \mathcal{T}^{5} \notag \\ 
&+ \Big(\big(9 \mathcal{C}^{2} -24 \mathcal{S}^{2} \big) |d_{1}|^{2} + \big(9 \mathcal{S}^{2} -24 \mathcal{C}^{2} \big) |d_{2}|^{2} 
+33 \mathcal{SC}  \big(d_{1},d_{2}\big)    \Big)  \mathcal{T}^{4}     \notag \\ 
&+ \Big(18 \mathcal{SC} \big( |d_{2}|^{2} - |d_{1}|^{2}\big)  
+9 \big(\mathcal{C}^{2} -\mathcal{S}^{2} \big) \big(d_{1},d_{2}\big)   \Big)  \mathcal{T}^{3}  \notag \\ 
&+ \Big(-9 \mathcal{SC}  \big(d_{1},d_{2}\big) 
+9 \big(\mathcal{S}^{2} |d_{1}|^{2} + \mathcal{C}^{2} |d_{2}|^{2} \big)  \Big)  \mathcal{T}^{2}  \Big]    , 
\label{swfg00}  
\end{align}
\begin{align}
\tau_{ij} = \frac{1}{8\pi G\eta^{8}} \delta_{ij}  \Big[& \Big( \big(3\mathcal{S}^{2}-2\mathcal{C}^{2}  \big) |d_{1}|^{2}  
+ \big( 3\mathcal{C}^{2}-2\mathcal{S}^{2}\big)  |d_{2}|^{2} -5\mathcal{SC}  \big(d_{1},d_{2}\big)   \Big) 
 \mathcal{T}^{6}   \notag \\
&+ \Big(24 \mathcal{SC}\big( |d_{1}|^{2}  -  |d_{2}|^{2} \big) - 12 \big( \mathcal{C}^{2} - \mathcal{S}^{2}\big)  
  \big(d_{1},d_{2}\big)   \Big)  \mathcal{T}^{5} \notag \\ 
&+  \Big( \big(15 \mathcal{C}^{2}-26\mathcal{S}^{2}  \big) |d_{1}|^{2}  - \big(26 \mathcal{C}^{2}-15\mathcal{S}^{2}  \big) |d_{2}|^{2}
+41 \mathcal{SC}  \big(d_{1},d_{2}\big)    \Big)  \mathcal{T}^{4}     \notag \\ 
&+ \Big(30 \mathcal{SC} \big( |d_{2}|^{2} - |d_{1}|^{2}\big)  
+15 \big(\mathcal{C}^{2} -\mathcal{S}^{2} \big) \big(d_{1},d_{2}\big)   \Big)  \mathcal{T}^{3}  \notag \\ 
&+ \Big(-19 \mathcal{SC}  \big(d_{1},d_{2}\big) - 4 \big(\mathcal{C}^{2} |d_{1}|^{2} + \mathcal{S}^{2} |d_{2}|^{2} \big)
+15 \big(\mathcal{S}^{2} |d_{1}|^{2} + \mathcal{C}^{2} |d_{2}|^{2} \big) \notag \\ 
&+ 4\eta^{2} \Big\{\mathcal{SC}  \big(d_{1},d_{2}\big) +  \big(\mathcal{C}^{2} |d_{1}|^{2} 
+ \mathcal{S}^{2} |d_{2}|^{2} \big)   \Big\}  \Big)  \mathcal{T}^{2}  \notag \\ 
&+ \Big( 8 \mathcal{SC} \big( |d_{1}|^{2} - |d_{2}|^{2}\big)  
+ 4 \big(\mathcal{S}^{2}-\mathcal{C}^{2}\big) \big(d_{1},d_{2}\big)  \notag \\ 
&+ 4\eta^{2} \Big\{ \big(\mathcal{C}^{2}-\mathcal{S}^{2} \big) \big(d_{1},d_{2}\big) 
 -2\mathcal{SC} \big(|d_{1}|^{2} - |d_{2}|^{2}  \big) \Big\}
 \Big)  \mathcal{T}    \notag \\ 
&+4 (1-\eta^{2}) \Big( -  \big(\mathcal{S}^{2} |d_{1}|^{2} + \mathcal{C}^{2} |d_{2}|^{2} \big)
+  \mathcal{SC} \big(d_{1},d_{2}\big)     \Big)    \Big]   . \label{swfgij} 
\end{align}

\subsection{Comoving gauge}

\subsubsection{MDE}

\begin{align}
\tau_{00} 
= & 
\frac{1}{8\pi G} \bigg[ 
\frac{77k^{2}}{9} |{c}_{1}|^{2}  
- \frac{77k^{2}}{9 \eta^{5}} \big({c}_{1},{c}_{2}\big)  
+\frac{77k^{2}}{9\eta^{10}} |{c}_{2}|^{2}  \bigg]  
\,  , 
\label{dustcg00}  
\\
\tau_{ij} 
= & 
\frac{1}{8\pi G} \delta_{ij} \bigg[
\frac{16}{9}|{c}_{1}|^{2}  
+\frac{13k^{2}}{27} |{c}_{1}|^{2}  -\frac{64}{9\eta^{2}} |{c}_{1}|^{2}  
+ \frac{48+13k^{2}}{27\eta^{5}} \big({c}_{1}, {c}_{2}\big)  
- \frac{64}{9\eta^{7}} \big({c}_{1},{c}_{2}\big) 
\nonumber \\
& \hspace{4em}
+ \frac{48+13k^{2}}{27\eta^{10}} |{c}_{2}|^{2}   
-\frac{64}{9\eta^{12}} |{c}_{2}|^{2}   \bigg] 
\,   . 
\label{dustcgij} 
\end{align}

\subsubsection{RDE in long-wavelength limit}

\begin{align}
\tau_{00} =  \frac{1}{8\pi G\eta^8} \Big[ &
\frac{45}{2} |d_{2}|^{2}  \mathcal{T}^{2}
+\frac{45}{2} |d_{2}|^{2}  \mathcal{T}^{4}
-\frac{15 \big(d_{1},d_{2}\big) }{2}  \mathcal{T}^{5}  
- 3 \big(d_{1},d_{2}\big)   \mathcal{T}^{7}
+\frac{5\big(|d_{1}|^{2} - |d_{2}|^{2} \big) }{2} \mathcal{T}^{8}   
 \nonumber \\
&
+ {\cal O} \Big( \mathcal{T}^{9} \Big)  \Big]    , \label{lwcg00}  \\
\tau_{ij} = \frac{1}{8\pi G} \delta_{ij} & \Big[ \Big( \frac{1}{\eta^{6}}- \frac{1}{\eta^{8}} \Big) |d_{2}|^{2}
+\Big( \frac{1}{\eta^{6}}+ \frac{1}{2\eta^{8}} \Big)|d_{2}|^{2}  \mathcal{T}^{2}
-\frac{1}{3}\Big( \frac{1}{\eta^{6}}- \frac{1}{\eta^{8}} \Big)  \big(d_{1},d_{2}\big)  \mathcal{T}^{3}    
\nonumber \\
&
+\frac{3}{2\eta^{8}}|d_{2}|^{2} \mathcal{T}^{4}     
- \frac{1}{15}\Big( \frac{2}{\eta^{6}}+ \frac{11}{2\eta^{8}} \Big)  \big(d_{1},d_{2}\big) 
 \mathcal{T}^{5}        
  +\frac{1}{9} \Big(  \frac{1}{\eta^{6}} - \frac{1}{\eta^{8}} \Big) 
  \big( |d_{1}|^{2} - |d_{2}|^{2} \big)
 \mathcal{T}^{6}    
 \nonumber \\
& 
 + \frac{1}{35} \Big( \frac{2}{\eta^{6}} - \frac{9}{\eta^{8}} \Big)   \big(d_{1},d_{2}\big) 
 \mathcal{T}^{7}   
 -\frac{1}{45} \Big( \frac{1}{\eta^{6}} - \frac{17}{2\eta^{8}} \Big) \big( |d_{1}|^{2} - |d_{2}|^{2} \big) 
  \mathcal{T}^{8}       
+ {\cal O} \Big( \mathcal{T}^{9} \Big)    \Big]    . \label{lwcgij} 
\end{align}

\subsubsection{RDE in short-wavelength limit}

\begin{align}
\tau_{00} = \frac{1}{8\pi G} \times \frac{45}{2\eta^{8}} \Big[ &
\Big( \mathcal{C}^{2} |d_{1}|^{2} + \mathcal{S}^{2} |d_{2}|^{2} 
+\mathcal{SC}  \big(d_{1},d_{2}\big)    \Big)  \mathcal{T}^{4}     \notag \\ 
& + \Big(2 \mathcal{SC} \big( |d_{2}|^{2} - |d_{1}|^{2}\big)  
+\big(\mathcal{C}^{2} -\mathcal{S}^{2} \big) \big(d_{1},d_{2}\big)   \Big)  \mathcal{T}^{3}  \notag \\ 
& + \Big( \mathcal{S}^{2} |d_{1}|^{2} + \mathcal{C}^{2} |d_{2}|^{2} 
- \mathcal{SC}  \big(d_{1},d_{2}\big)  \Big)  \mathcal{T}^{2}  \Big]    , \label{swcg00}   
\end{align}
\begin{align}
\tau_{ij} = \frac{1}{8\pi G} \delta_{ij}  \Big[& \frac{3}{2\eta^{8}} \Big( \mathcal{C}^{2} |d_{1}|^{2}+ \mathcal{S}^{2} |d_{2}|^{2} +\mathcal{SC}  \big(d_{1},d_{2}\big)   \Big) 
 \mathcal{T}^{4}   \notag \\
& +\frac{3}{2\eta^{8}} \Big(2 \mathcal{SC} \big( |d_{2}|^{2} - |d_{1}|^{2}\big)  
+ \big(\mathcal{C}^{2} -\mathcal{S}^{2} \big) \big(d_{1},d_{2}\big)   \Big)  \mathcal{T}^{3}  \notag \\ 
& + \Big( \frac{1}{2\eta^{8}} \Big\{ -5 \mathcal{SC}  \big(d_{1},d_{2}\big) - 2 \big(\mathcal{C}^{2} |d_{1}|^{2} + \mathcal{S}^{2} |d_{2}|^{2} \big)
+3 \big(\mathcal{S}^{2} |d_{1}|^{2} + \mathcal{C}^{2} |d_{2}|^{2} \big) \Big\} \notag \\ 
& + \frac{1}{\eta^{6}} \Big\{\mathcal{SC}  \big(d_{1},d_{2}\big) +  \big(\mathcal{C}^{2} |d_{1}|^{2} 
+ \mathcal{S}^{2} |d_{2}|^{2} \big)   \Big\}  \Big)  \mathcal{T}^{2}  \notag \\ 
& + \Big(  \frac{1}{\eta^{8}} \Big\{ 2\mathcal{SC} \big( |d_{1}|^{2} - |d_{2}|^{2}\big)  
+ \big(\mathcal{S}^{2}-\mathcal{C}^{2}\big) \big(d_{1},d_{2}\big) \Big\}  \notag \\ 
& + \frac{1}{\eta^{6}} \Big\{ \big(\mathcal{C}^{2}-\mathcal{S}^{2} \big) \big(d_{1},d_{2}\big) 
 -2\mathcal{SC} \big(|d_{1}|^{2} - |d_{2}|^{2}  \big) \Big\}
 \Big)  \mathcal{T}    \notag \\ 
& +\Big( \frac{1}{\eta^{8}} -\frac{1}{\eta^{6}} \Big) \Big( -  \big(\mathcal{S}^{2} |d_{1}|^{2} + \mathcal{C}^{2} |d_{2}|^{2} \big)
+  \mathcal{SC} \big(d_{1},d_{2}\big)     \Big)    \Big]   . \label{swcgij} 
\end{align}

\end{document}